\documentclass[10pt,twoside,a4paper]{article}
\usepackage{units}

\usepackage[a4paper,centering,body={31pc,49pc}]{geometry}

\usepackage[numbers,sort,merge]{natbib}

\usepackage{multicol}

\usepackage[toc,page]{appendix}

\usepackage{epsfig}
\usepackage{subfigure}

\usepackage{amsmath}
\usepackage{amsfonts}
\usepackage{amssymb}
\usepackage{amsthm}

\usepackage{fancyhdr}
\usepackage{hyperref}

\usepackage[T1]{fontenc}\usepackage{txfonts}\usepackage[full]{textcomp}

\renewcommand{\Re}{\mathrm{Re}}%

\DeclareMathOperator{\cosec}{cosec}
\providecommand{\Ent}[1]{\lfloor #1 \rfloor}

\numberwithin{equation}{section}

\renewcommand{\sectionmark}[1]%
 {\markboth{\thesection:\ #1}{}}

\begin{document}

\thispagestyle{empty}


\title{\bfseries Convergence Analysis of the Summation \\
  of the Euler Series by \\
  Pad\'{e} Approximants and the Delta Transformation}

\author{Riccardo Borghi$^1$ and Ernst Joachim Weniger$^2$ \\[2.5\jot]
  $^1$ Dipartimento di Ingegneria, \\
  Universit\`{a} ``Roma Tre'' , I-00144 Rome, Italy \\
  $^2$ Instit{u}t f\"{u}r Physikalische und Theoretische Chemie, \\
  Universit\"{a}t Regensburg, D-93040 Regensburg, Germany}

\date{Submitted to \texttt{arXiv.org}: 10 May 2014}

\maketitle

\begin{abstract}
  Sequence transformations are valuable numerical tools that have been
  used with considerable success for the acceleration of convergence and
  the summation of diverging series.  However, our understanding of their
  theoretical properties is far from satisfactory. The Euler series
  $\mathcal{E}(z) \sim \sum_{n=0}^{\infty} (-1)^{n} n! z^{n}$ is a very
  important model for the ubiquitous factorially divergent perturbation
  expansions in physics. In this article, we analyze the summation of the
  Euler series by Pad\'{e} approximants and the delta transformation
  [E. J. Weniger, Comput. Phys. Rep. \textbf{10}, 189 (1989), Eq.\
  (8.4-4)] which is a powerful nonlinear Levin-type transformation that
  works very well in the case of strictly alternating convergent or
  divergent series.  Our analysis is based on a new factorial series
  representation of the truncation error of the Euler series~[R. Borghi,
  Appl. Num. Math. \textbf{60}, 1242 (2010)]. We derive explicit
  expressions for the transformation errors of Pad\'{e} approximants and
  of the delta transformation.  A subsequent asymptotic analysis proves
  \emph{rigorously} the convergence of both Pad\'e and delta. Our
  asymptotic estimates clearly show the superiority of the delta
  transformation over Pad\'{e}. This is in agreement with previous
  numerical results.
\end{abstract}

\begin{quote}

  {\bfseries AMS classification scheme numbers:} 
   26C15, 
    33-XX, 
     40A05, 
      41A20, 
       41A21 
     
\end{quote}

\newpage

\tableofcontents

\newpage


%
\typeout{==> Section: Introduction}
\section{Introduction}
\label{Sec:Introduction}
%

In mathematics, divergent series have been a highly controversial topic
\cite{Burkhardt/1911,Ferraro/2008,Tucciarone/1973}, and to some extend
they still are (see for example \cite[Appendix D]{Weniger/2008} and
references therein). Already Euler had frequently used divergent series,
albeit not necessarily in a sufficiently rigorous way
\cite{Barbeau/1979,Barbeau/Leah/1976,Digernes/Varadarajan/2010,Dutka/1996,%
  Kozlov/2007,Varadarajan/2007}. But as soon as the concept of
convergence had been better understood, Euler's approach was criticized,
and in the earlier part of the nineteenth century a strong tendency
emerged to ban divergent series completely from rigorous mathematics.

Mathematical orthodoxy prevailed for a while, but at the end of the
nineteenth century it had become clear that divergent series were simply
too useful to be neglected. In addition, the work of mathematicians like
Borel, Pad\'{e}, Poincar\'{e}, Stieltjes, and others ultimately led in
the later part of the nineteenth century to a mathematically rigorous
theory of divergent series (for more details and additional references,
see for example the book by Kline \cite[Chapter 47]{Kline/1972}, or the
articles by Ferraro \cite{Ferraro/1999} and Tucciarone
\cite{Tucciarone/1973}).

The work of the mathematicians mentioned above also showed that divergent
series can be used for computational purposes if they are combined with
suitable summation techniques. This observation turned out to be
extremely consequential since it later inspired a considerable amount of
research on summation techniques. As a contemporary example of this
research, see the proceedings of a recent conference published in
\cite{Brezinski/RedivoZaglia/Weniger/2010a,Weniger/2010a}.

Rigorous convergence proofs of summation techniques are notoriously
difficult, in particular in the case of nonlinear techniques like the
ones considered in this article. This explains why relatively little had
been achieved so far. However, physicist normally do not hesitate to
employ numerical techniques in general and summation techniques in
special, even if their theoretical properties are not fully
understood. They are usually satisfied with convincing numerical evidence
that a given technique produces correct results at least for some
important special cases.

Generally speaking, the practical usefulness of divergent series in
physics has always been more important than concerns about mathematical
rigor. A classical example is Stirling's series for the logarithm of the
gamma function \cite[Eq.\
(5.11.1)]{Olver/Lozier/Boisvert/Clark/2010}. Stirling's series diverges,
but is \emph{semiconvergent} (for a condensed review of the concept of
semiconvergence, which had been introduced by Stieltjes
\cite{Stieltjes/1886} already in 1886, see \cite[Appendix
E]{Weniger/2012}). Thus, the truncation of Stirling's divergent series
produces excellent approximations for large arguments. Those approximants 
played a major role in the formal development of statistical mechanics. Another
example is Poincar\'{e}'s seminal work on asymptotic series
\cite{Poincare/1886}, which was inspired by his work in astronomy where
truncated asymptotic series turned out to be extremely useful.

In quantum physics, divergent series are indispensable. Already in 1952,
Dyson \cite{Dyson/1952} had argued that perturbation expansions in
quantum electrodynamics must diverge factorially. Around 1970, Bender and
Wu \cite{Bender/Wu/1969,Bender/Wu/1971,Bender/Wu/1973} showed in their
pioneering work on anharmonic oscillators that factorially divergent
perturbation expansions occur also in nonrelativistic quantum
mechanics. Later, it was found that factorially divergent perturbation
expansions are actually the rule in quantum physics rather than the
exception (see for example the articles by Fischer \cite[Table
1]{Fischer/1997} and Suslov \cite{Suslov/2005}, or the articles reprinted
in the book by Le Guillou and Zinn-Justin
\cite{LeGuillou/Zinn-Justin/1990}). The observation, that the
perturbation expansions of quantum physics diverge almost by default,
caused a renaissance of divergent series in theoretical physics.

From the perspective of a theoretical physicist, a divergent series is a
fully acceptable mathematical and computational tool, provided that this
series is summable to a \emph{finite} generalized limit. Of course, there
remains the practically very important question how such a divergent
series can be summed in an \emph{effective} and numerically
\emph{reliable} way.

The most important summation techniques used by theoretical physicists
are Borel summation \cite{Borel/1899}, which replaces a divergent
perturbation expansion by a Laplace-type integral, and Pad\'{e}
approximants \cite{Pade/1892}, which transform the partial sums of a
(formal) power series to rational functions. There is an extensive
literature on these summation techniques and their applications not only
in theoretical physics. Borel summation is for example discussed in books
by Costin \cite{Costin/2009}, Shawyer and Watson
\cite{Shawyer/Watson/1994}, and Sternin and Shatalov
\cite{Sternin/Shatalov/1996}, and the most complete treatment of Pad\'{e}
approximants can be found in the monograph by Baker and Graves-Morris
\cite{Baker/Graves-Morris/1996}. Both Borel summation and Pad\'{e}
approximants have been remarkably successful, and there are far too many
articles describing their applications to be cited here. Nevertheless,
these summation methods have -- like all other numerical techniques --
certain shortcomings and limitations. Therefore, it is certainly
justified to look for alternatives that are at least in some cases
capable of producing better results.

In recent years, a lot of research has been done on so-called sequence
transformations which turned out to be very useful numerical tools for
the summation of divergent series. Of course, there is also an extended
recent literature, in particular on \emph{non-linear} and
\emph{non-regular} sequence transformations. Examples are the monographs
by Brezinski \cite{Brezinski/1977,Brezinski/1978}, Brezinski and Redivo
Zaglia \cite{Brezinski/RedivoZaglia/1991a}, Sidi \cite{Sidi/2003}, and
Wimp \cite{Wimp/1981}, the review by Weniger \cite{Weniger/1989}, or the
proceedings \cite{Brezinski/RedivoZaglia/Weniger/2010a,Weniger/2010a} of
a recent conference. 

The majority of these references emphasize the theoretical properties of
sequences transformations and are mainly of interest for mathematicians
who want to work \emph{on} sequence transformations, and not so much for
physicists who usually want to work \emph{with} sequence
transformations. But recently, the situation has changed and several
books have appeared that describe how nonlinear sequence transformations
can be employed effectively as computational tools. Examples are books by
Bornemann, Laurie, Wagon, and Waldvogel \cite[Appendix
A]{Bornemann/Laurie/Wagon/Waldvogel/2004}, by Gil, Segura, and Temme
\cite[Chapter 9]{Gil/Segura/Temme/2007} (compare also the related review
articles by Gil, Segura, and Temme \cite{Gil/Segura/Temme/2011} and by
Temme \cite{Temme/2007}), and the last (3rd) edition of \emph{Numerical
  Recipes} \cite[Chapter 5]{Press/Teukolsky/Vetterling/Flannery/2007}
(Weniger's detailed review \cite{Weniger/2007d} of the treatment of
sequence transformations in \emph{Numerical Recipes}
\cite{Press/Teukolsky/Vetterling/Flannery/2007} can be downloaded from
the Numerical Recipes web page). Last, but not least: Sequence
transformations are also treated in the recently published NIST Handbook
of Mathematical Functions \cite[Chapter 3.9 Acceleration of
Convergence]{Olver/Lozier/Boisvert/Clark/2010} and in a very recent book
by Trefethen \cite[Chapter 28]{Trefethen/2013}.

In this context, it may be of interest to note that Pad\'{e}
approximants, which transform the partial sums of a (formal) power series
to a doubly indexed sequence of rational functions, can also be viewed to
be nothing but a special class of sequence transformations.

Levin-type transformations \cite{Levin/1973,Weniger/1989,Weniger/2004},
whose most important features are reviewed in Section
\ref{Sec:Levin-TypeTransformations}, were found to be a remarkably
effective and powerful class of sequence transformations. They differ
substantially from other, better known sequence transformations as for
example Wynn's epsilon algorithm \cite{Wynn/1956a}, which will be
discussed in Section \ref{Sub:WynnEspAl}. They use as input data not only
finite substrings of a sequence $\{ s_{n} \}_{n=0}^{\infty}$, whose
elements may be the partial sums $s_{n} = \sum_{k=0}^{n} a_{k}$ of a
slowly convergent or divergent infinite series, but also finite
substrings of a sequence $\{ \omega_{n} \}_{n=0}^{\infty}$ of explicit
estimates of the truncation errors of $\{ s_{n} \}_{n=0}^{\infty}$. The
explicit incorporation of the information contained in the remainder
estimates $\{ \omega_{n} \}_{n=0}^{\infty}$ into the transformation
process is to a large extend responsible for the frequently superior
performance of Levin-type transformations. As discussed later in more
detail, numerous successful applications of Levin-type transformations
are described in the literature, and they are now also discussed in the
recently published NIST Handbook \cite[Chapter 3.9(v) Levin's and
Weniger's Transformations]{Olver/Lozier/Boisvert/Clark/2010}.

From the perspective of somebody who only wants to employ Levin-type
transformations as computational tools, the situation looks quite
good. However, a numerical technique is only fully satisfactory if it is
augmented by a sufficiently detailed and easily applicable convergence
theory. In that respect, the situation is not good at all. Our
understanding of the convergence properties of Levin-type transformations
is far from satisfactory, and it is particularly bad when it comes to the
summation of factorially divergent series.

In the case of Pad\'{e} approximants, the situation is much better. If
the input data are the partial sums of a Stieltjes series, whose most
important properties are reviewed in Appendix \ref{App:StieFunStieSer},
it can be proved rigorously that certain subsequences of the Pad\'{e}
table converge to the value of the corresponding Stieltjes function
(compare our discussion in Section \ref{Sub:PadeApprox}). This is very
convenient. If we can prove that a factorially divergent power series is
a Stieltjes series and that it also satisfies the Carleman condition
(\ref{Carleman_Condition}), we immediately know that this series is
Pad\'{e} summable.

Because of the use of explicit remainder estimates and the exploitation
of the information contained in them, Levin-type transformations often
produce remarkably good convergence acceleration and summation results.
But when it comes to a theoretical analysis of the properties of
Levin-type transformations, remainder estimates cause additional
complications that make rigorous proofs more difficult. Because of the
specific features of Levin-type transformations, we can only hope to
prove that a \emph{certain} Levin-type transformation produces convergent
results when applied to the elements of a \emph{given} slowly convergent
or divergent input sequence $\{ s_{n} \}_{n=0}^{\infty}$. In this
article, we only consider the summation of the factorially divergent
\emph{Euler series} (\ref{Def:EulerSer}), which is -- as discussed in
more detail in Section \ref{Sec:EulerSeries} -- the simplest prototype of
the ubiquitous factorially divergent series expansions.

It will become clear later that the technical problems, which occur in
our convergence studies, differ substantially among the various
Levin-type transformations. In this article, we only consider the
so-called delta transformation \cite[Eq.\ (8.4-4)]{Weniger/1989}.  This
transformation is known to be very effective in the case of both
convergent and divergent alternating series. It sums the Euler series or
related factorially divergent expansions much more effectively than for
example Pad\'{e} approximants. It will become clear later that this delta
transformation is particularly suited for a theoretical convergence
analysis of the summation of the Euler series.

The Euler series (\ref{Def:EulerSer}) is a Stieltjes series. Therefore,
it is guaranteed that its partial sums are Pad\'{e} summable to the Euler
integral (\ref{Def:EulerIntegral}). It is, however, not so well known
that Pad\'{e} approximants to the Euler series can be expressed in closed
form with the help of Drummond's sequence transformation
\cite{Drummond/1972} which is another Levin-type transformation
\cite[Sections 9.5 and 13.2]{Weniger/1989}. The resulting expression for
this Pad\'{e} approximant \cite[Eq.\ (13.3-5)]{Weniger/1989} can
according to Eq.\ (\ref{Pade.0}) be brought into a form which resembles
the finite difference operator representation
(\ref{Def:DeltaTasformation}) of the delta transformation. Accordingly,
our convergence analysis of the delta summation of the Euler series can
easily be modified to cover the Pad\'{e} summation of the Euler series.
In this way, we were able to derive closed-form expressions for the
transformation errors of both the Pad\'e and the delta summation of the
Euler series. Our subsequent asymptotic analysis proves rigorously that
both summation processes converge. In addition, we obtain estimates for
the rate of convergence of both processes, which are in agreement with
previous numerical results and which demonstrate the superiority of the
delta transformation.

Our article is organized as follows: In Section \ref{Sec:EulerSeries}, we
briefly review the Euler series. In Section \ref{Sub:PadeApprox}, the for
our purposes most important properties of Pad\'{e} approximants are
reviewed, and in Section \ref{Sub:WynnEspAl} we briefly review Wynn's
epsilon algorithm. Section \ref{Sec:Levin-TypeTransformations} discusses
the special features of Levin-type transformations. This completes the
introductory part which should enable nonspecialist readers to appreciate
the core of our work. This is contained in Sections
\ref{Sec:SummEulSerDeltaTrans} - \ref{Sec:NumericalExamples}. Finally, we
present in Section \ref{Sec:Conclusions} some conclusions and some
outlook on possible extensions of our work.


%
\typeout{==> Section: The Euler Series as a Paradigm for Factorial
  Divergence}
\section{The Euler Series as a Paradigm for Factorial Divergence}
\label{Sec:EulerSeries}
%

The classic example of a factorially divergent power series is the
so-called \emph{Euler series}
\begin{equation}
  \label{Def:EulerSer}
  \mathcal{E}_{} (z) \; \sim \; 
  \sum_{m=0}^{\infty} \, (-1)^{m} \, m! \, z^{m} 
  \; = \; {}_{2} F_{0} (1, 1; -z) \, ,
  \qquad z \to 0\, ,
\end{equation}
whose summation and interpretation had already been studied by Euler (see
for example \cite[pp.\ 323 - 324]{Bromwich/1991}, \cite[pp.\ 26 -
29]{Hardy/1949} or \cite{Barbeau/1979,Barbeau/Leah/1976}). If $\vert
\arg(z) \vert < \pi$, the Euler series is asymptotic as $z \to 0$ in the
sense of Poincar\'{e} \cite{Poincare/1886} to the so-called \emph{Euler
  integral}
\begin{equation}
  \label{Def:EulerIntegral}
  \displaystyle
  \mathcal{E}(z) \; = \; \int_{0}^{\infty} \, 
  \frac{\exp (-t)}{1 + z t} \, \mathrm{d} t \, ,
  \qquad \vert \arg (z) \vert < \pi \, .
\end{equation}
If we set $\Phi (t) = 1 - \exp (-t)$ in Eqs.\ (\ref{Def_StieFun}),
(\ref{Def_StieMom}), and (\ref{Def_StieSer}), we immediately see that the
Euler series (\ref{Def:EulerSer}) is a Stieltjes series and that the
associated Euler integral (\ref{Def:EulerIntegral}) is a Stieltjes
function. It also follows from Eq.\ (\ref{Def_StieFunParSum}) that the
partial sum
\begin{equation}
  \label{Def:ParSunEuSer}
  \mathcal{E}_{n} (z) \; = \; \sum_{\nu=0}^{n} \, 
  (-1)^{\nu} \, \nu! \, z^{\nu} \, , \qquad n \in \mathbb{N}_{0} \, ,
\end{equation}
of the Euler series can according to
\begin{equation}
  \label{ParSumEuFun=EuInt+TruncErr}
  \mathcal{E}_{n} (z) \; = \; \mathcal{E} (z) \, + \,  
   \mathcal{R}_{n} (\mathcal{E}; z) \, , \qquad n \in \mathbb{N}_{0} \, ,
   \quad \vert \arg (z) \vert < \pi \, ,
\end{equation}
be expressed by the Euler integral plus a truncation error term 
\begin{equation}
  \label{Def:TruncErrEuSer}
  \mathcal{R}_{n} (\mathcal{E}; z) \; = \; - (-z)^{n+1} \, 
   \int_{0}^{\infty} \, \frac{t^{n+1}\exp (-t) \mathrm{d} t} {1+zt} 
    \, , \qquad n \in \mathbb{N}_{0} \, ,
   \quad \vert \arg (z) \vert < \pi \, .
\end{equation}
The integral in Eq.\ (\ref{Def:TruncErrEuSer}) is also a Stieltjes
integral.  Equation (\ref{StieSerErrEst}) implies that the truncation
error term (\ref{Def:TruncErrEuSer}) is bounded in magnitude by the first
term neglected in the partial sum (\ref{Def:ParSunEuSer}):
\begin{equation}
  \label{EuSerTruncErrBound}
  \bigl\vert \mathcal{R}_{n} (\mathcal{E}; z) \bigr\vert \; \le \;
  \begin{cases}
    (n+1)! \, \vert z^{n+1} \vert \, ,
    & \qquad \vert \arg (z) \vert \le \pi /2 \, , 
    \rule[-2ex]{0cm}{1em} \\
    (n+1)! \, \vert z^{n+1} \cosec \bigl( \arg (z) \bigr) \vert \, , 
   & \qquad \pi /
    2 < \vert \arg (z) \vert < \pi \, .
  \end{cases}  
\end{equation}

The terms of the partial sum (\ref{Def:ParSunEuSer}) decrease with
increasing index $n$ as long as $(n+1)\vert z \vert < 1$ holds. Thus, for
$z > 0$ the Euler series should be truncated at $n \approx
(1-z)/z$. Accordingly, the accuracy, which can be obtained by optimally
truncating the Euler series at its minimal term, depends strongly on the
magnitude of the argument $z$ (see for example \cite[Section
2]{Ferreiraa/Lopez/Mainar/Temme/2005}). The inclusion of higher terms
only leads to a deterioration of accuracy, and ultimately the Euler
series diverges for every nonzero $z \in \mathbb{C}$. This observation
confirms once more that the radius of convergence of the Euler series
${}_2 F_0 (1,1;- z)$ is zero.

Factorially divergent \emph{inverse} power series occur abundantly in
special function theory. By means of some elementary operations, it can
be shown that the Euler integral can be expressed as an exponential
integral $E_{1}$ with argument $1/z$ \cite[Eq.\
(6.2.2)]{Olver/Lozier/Boisvert/Clark/2010}:
\begin{equation}
  \label{EuInt<->ExoInt}
  \mathcal{E} (z) \; = \; \frac{\exp(1/z)}{z} \, \int_{z}^{\infty} \, 
  \frac{\exp (-t)}{t} \, \mathrm{d} t \; = \; 
  \frac{\exp(1/z)}{z} \, E_{1} (1/z) \, .
\end{equation}
Accordingly, the Euler series with argument $1/z$ corresponds to the
asymptotic expansion of the exponential integral as $z \to \infty$
\cite[Eq.\ (6.12.1)]{Olver/Lozier/Boisvert/Clark/2010}:
\begin{equation}
  \label{ExpInt_AsySer}
  z \, \exp(z) \, E_{1} (z) \; \sim \;
  \sum_{m=0}^{\infty} \, \frac{(-1)^m m!}{z^m}
  \; = \; {}_2 F_0 (1, 1; -1/z) \, , \qquad z \to \infty \, ,
  \quad \vert \arg (z) \vert < 3\pi/2 \, .
\end{equation}

Asymptotic series involving a factorially divergent generalized
hypergeometric series ${}_2 F_0$ occur also among other special
functions. Examples are the asymptotic expansion of the modified Bessel
function of the second kind \cite[Eq.\
(10.40.2)]{Olver/Lozier/Boisvert/Clark/2010},
\begin{align}
  \label{BesK_AsySer}
  K_{\nu} (z) \; \sim \; &
  [\pi / (2 z)]^{1/2} \, \exp(- z) \,
  {}_2 F_0 \bigl( 1/2 + \nu, 1/2 - \nu; - 1/(2 z) \bigr) \, ,
  \notag
  \\
   & \qquad \vert z \vert \to \infty \, , 
    \qquad \vert \arg (z) \vert < 3\pi/2 \, ,
\end{align}
or the asymptotic expansion of the Whittaker function of the second kind
\cite[Eq.\ (13.19.3)]{Olver/Lozier/Boisvert/Clark/2010},
\begin{align}
  \label{WhittW_AsySer}
  W_{\kappa,\mu} (z) \; \sim \; & \exp(-z/2) \, z^{\kappa} \, {}_2 F_0
  \bigl( \mu - \kappa + 1/2, - \mu - \kappa + 1/2; - 1/z \bigr) \, ,
  \notag
  \\
   & \qquad \vert z \vert \to \infty \, , 
    \qquad \vert \arg (z) \vert < 3\pi/2 \, .
\end{align}

These expansions have essentially the same features as the Euler series
(\ref{Def:EulerSer}) or rather the asymptotic series
(\ref{ExpInt_AsySer}) for $E_1 (z)$. By optimally truncating them in the
vicinity of their minimal terms, approximations to the special functions
can be obtained whose accuracies depend on the magnitude of the effective
argument $1/z$. However, these asymptotic expansions diverge factorially
for every $\vert z \vert < \infty$ if additional terms beyond the minimal
term are included.

For \emph{very} large arguments, suitably truncated partial sums of the
divergent asymptotic inverse power series mentioned above provide
sufficiently accurate approximations to the special functions they
represent. These approximations can be extremely useful numerically, but
it can also happen that these approximations are already for only
moderately large arguments too inaccurate to be practically useful.

However, the partial sums of such a divergent inverse power series can be
used to construct either Pad\'{e} approximants or other rational
approximants based on sequence transformations (see for example
\cite[Section VI]{Weniger/2004}). These rational approximants are often
able to provide remarkably accurate results even for relatively small
arguments. In \cite[Section 13.3]{Weniger/1989} it was shown that the
transformations $d_{k}^{(n)} (\beta, s_{n})$ and $\delta_{k}^{(n)}
(\beta, s_{n})$ defined in Eqs.\ (\ref{dLevTr}) and (\ref{dWenTr}),
respectively, sum the divergent asymptotic inverse power series
(\ref{ExpInt_AsySer}) for the exponential integral $E_{1} (z)$ much more
effectively than Pad\'{e} approximants. Qualitatively similar summation
results were also obtained in the case of divergent asymptotic inverse
power series (\ref{BesK_AsySer}) for the modified Bessel function
$K_{\nu} (z)$ \cite{Weniger/Cizek/1990}, and in the case of divergent
asymptotic inverse power series (\ref{WhittW_AsySer}) for the Whittaker
function $W_{\kappa,\mu} (z)$ \cite{Weniger/1996d}.


%
\typeout{==> Section: Pad\'{e} Approximation and Wynn's Epsilon Algorithm}
\section{Pad\'{e} Approximation and Wynn's Epsilon Algorithm}
\label{Sec:PadeApproxWynnEspAl}
%

%
\typeout{==> Subsection: A Review of Pad\'{e} Approximation}
\subsection{A Review of Pad\'{e} Approximation}
\label{Sub:PadeApprox}
%

The rational approximants, which are named after Pad\'{e}
\cite{Pade/1892} although they are actually much older (see for example
\cite[Chapters 4.5, 5.2.5, and 6.4]{Brezinski/1991a} or the more compact
treatment in \cite[Section 2]{Brezinski/1996}), are extremely important
computational tools. There are far too many books as well as countless
articles describing work both \emph{on} and \emph{with} Pad\'{e}
approximants to be cited in this article. Our major source is the
monograph by Baker and Graves-Morris \cite{Baker/Graves-Morris/1996} --
occasionally nicknamed \emph{Red Bible} -- which extends the older book
by Baker \cite{Baker/1975} and which provides the currently most detailed
treatment of Pad\'{e} approximants. Of course,
\cite{Baker/Graves-Morris/1996} also contains a wealth of useful
references.

In this Section, we only review those algebraic and convergence
properties of Pad\'{e} approximants that are later needed to understand
the summation of the factorially divergent Euler series
(\ref{Def:EulerSer}). We also try to explain comprehensively in which
respect Pad\'{e} approximants resemble Levin-type transformations and how
they differ.

Let us assume that a function $f \colon \mathbb{C} \to \mathbb{C}$
possesses a (formal) power series
\begin{equation}
  \label{Def_PowSer}
  f (z) \; = \; \sum_{n=0}^{\infty} \, \gamma_{n} \, z^{n} \, ,
\end{equation}
which may or may not converge, and that
\begin{equation}
  \label{ParSum_PowSer}
  f_{n} (z) \; = \; \sum_{\nu=0}^{n} \, \gamma_{\nu} \, z^{\nu} \, ,
  \qquad n \in \mathbb{N}_{0} \, ,
\end{equation}
stands for a partial sum of this power series. 

The \emph{Pad\'{e} approximant} $[m/n]_{f} (z)$ to $f (z)$ is the ratio
of two polynomials $P^{[m/n]} (z)$ and $Q^{[m/n]} (z)$ of degrees $m$ and
$n$ in $z$:
\begin{subequations}
  \label{Def_PadeApprox}
  \begin{align}
    \label{Def_PadeApprox_a}
    [m/n]_{f} (z) & \; = \; \frac{P^{[m/n]} (z)}{Q^{[m/n]} (z)} \, ,
    \qquad m, n \in \mathbb{N}_{0} \, ,
    \\
    \label{Def_PadeApprox_b}
    P^{[m/n]} (z) & \; = \; p_{0} + p_{1} z + \dots + p_{m} z^{m} \; = \;
    \sum_{\mu=0}^{m} \, p_{\mu} z^{\mu} \, ,
    \\
    \label{Def_PadeApprox_c}
    Q^{[m/n]} (z) & \; = \; q_{0} + q_{1} z + \dots + q_{m} z^{n} \; = \;
    \sum_{\nu=0}^{n} \, q_{\nu} z^{\nu} \, .
  \end{align}
\end{subequations}
Only $m+n+1$ of the $m+n+2$ unspecified polynomial coefficients in Eq.\
(\ref{Def_PadeApprox}) are independent since multiplication of $P^{[m/n]}
(z)$ and $Q^{[m/n]} (z)$ by a common nonzero constant does not change the
value of the ratio $P^{[m/n]} (z)/Q^{[m/n]} (z)$. Normally, one chooses
$q_{0} = 1$ -- the so-called Baker condition -- in order to remove this
indeterminacy and to guarantee the analyticity of $[m/n]_{f} (z)$ at
$z=0$. The remaining $m+n+1$ unspecified coefficients $p_{0}$, $p_{1}$,
$\dots$, $p_{m}$ and $q_{1}$, $q_{2}$, $\dots$, $q_{n}$ are determined
via the requirement that the Maclaurin series of $[m/n]_{f} (z)$ agrees
with the power series (\ref{Def_PowSer}) for $f (z)$ as far as possible:
\begin{equation}
  \label{Acc_Thr_Ord_1}
  f (z) \, - \, \frac{P^{[m/n]} (z)}{Q^{[m/n]} (z)} \; = \; 
  \mathrm{O} \bigl( z^{m+n+1} \bigr) \, , \qquad z \to 0 \, .
\end{equation}
This \emph{accuracy-through-order} relationship or rather the equivalent
condition
\begin{equation}
  \label{Acc_Thr_Ord_2}
  Q^{[m/n]} (z) \, f (z) \, - \, P^{[m/n]} (z) \; = \; 
  \mathrm{O} \bigl( z^{m+n+1} \bigr) \, , \qquad z \to 0 \, .
\end{equation}
leads to a system of $l+m+1$ linear equations for the polynomial
coefficients \cite[Eqs.\ (1.5) and (1.7) on pp.\ 2 and
3]{Baker/Graves-Morris/1996}. If this system of equations possesses a
solution, which will be tacitly assumed henceforth, it follows from
Cramer's rule that a Pad\'{e} approximant can be defined by a ratio of
two determinants (see for example \cite[Eq.\ (1.27)]{Baker/1975}):
\begin{align}
  \label{PadeDetRep}
  [ m / n ]_f \, (z) \; = \; \frac {\begin{vmatrix} 
      \gamma_{m-n+1} & \gamma_{m-n+2} & \ldots & \gamma_{m}
      \\
      \gamma_{m-n+2} & \gamma_{m-n+3} & \ldots & \gamma_{m+1}
      \\
      \vdots & \vdots & \ddots \\ 
      \gamma_{m} & \gamma_{m+1} & \ldots & \gamma_{m+n}
      \\
      {\displaystyle \sum_{j=n}^{m} \gamma_{j-n} z^j} & {\displaystyle
        \sum_{j=n-1}^{m} \gamma_{j-n+1} z^j} & \ldots &
      {\displaystyle \sum_{j=0}^{m} \gamma_{j} z^j} \\
    \end{vmatrix}}
  {\begin{vmatrix} \gamma_{m-n+1} & \gamma_{m-n+2} & \ldots & \gamma_{m}
      \\
      \gamma_{m-n+2} & \gamma_{m-n+3} & \ldots & \gamma_{m+1}
      \\
      \vdots & \vdots & \ddots \\
      \gamma_{m} & \gamma_{m+1} & \ldots & \gamma_{m+n}
      \\
      z^n & z^{n-1} & \ldots & 1
    \end{vmatrix}} \, .
\end{align}
To the best of our knowledge, this determinantal expression was first
derived in 1881 by Frobenius \cite[Eqs.\ (5) and (6)]{Frobenius/1881}.

For numerical purposes, neither systems of coupled linear equations nor
the determinantal expression (\ref{PadeDetRep}) are particularly
appealing. Fortunately, numerous other more effective computational
algorithms are known (see for example \cite[Chapter
II.3]{Cuyt/Wuytack/1987}). The probably most important and also most
convenient recursive scheme for Pad\'{e} approximants is Wynn's epsilon
which will be discussed in Section \ref{Sub:WynnEspAl}.

There is a highly developed \emph{general} convergence theory of Pad\'{e}
approximants \cite[Chapter 6]{Baker/Graves-Morris/1996}). But in this
article, we concentrate on the convergence of Pad\'{e} approximants of
Stieltjes series satisfying Eqs.\ (\ref{Def_StieMom}) and
(\ref{Def_StieSer}) to their corresponding Stieltjes functions satisfying
Eq.\ (\ref{Def_StieFun}). This theory is summarized below in a highly
condensed way.

The poles of the Pad\'{e} approximants $[m+j/m]_{f} (z)$ with $j \ge -1$
to a Stieltjes series of the type of Eq.\ (\ref{Def_StieSer}) are simple,
lie on the negative real semi-axis, and have positive residues
\cite[Theorem 5.2.1]{Baker/Graves-Morris/1996}. Thus, it is guaranteed
that the poles of $[m+j/m]_{f} (z)$ mimic the cut of the corresponding
Stieltjes function along the negative real semi-axis.

The Euler series is a Stieltjes series with a \emph{zero} radius of
convergence.  In such a case, there is a dual problem if we try to
retrieve the corresponding Stieltjes function $f(z)$ from its divergent
series with the help of a summation technique. Firstly, a Maclaurin
series does not necessarily determine the corresponding function
uniquely. Just consider the function $g (z) = \exp (-1/z)$ whose
derivatives at $z=0$ all vanish. This implies that $f (z)$ satisfying
Eq.\ (\ref{Def_PowSer}) and $f (z) + g (z)$ possess the same Maclaurin
expansion. Therefore, we need a criterion that rules out the occurrence
of a function of the type of $g (z)$.

Secondly, the uniqueness of the integral representation
(\ref{Def_StieFun}) -- or equivalently the determinacy of the
corresponding Stieltjes moment problem -- is not guaranteed if the
Stieltjes moments (\ref{Def_StieMom}) increase too rapidly with
increasing index \cite{Graffi/Grecchi/1978}. Fortunately, there is a
comparatively simple \emph{sufficient} condition due to Carleman
\cite{Carleman/1926,*Carleman/1975} that solves these problems. If the
Stieltjes moments $\{ \mu_{n} \}_{n=0}^{\infty}$ defined by Eq.\
(\ref{Def_StieMom}) satisfy the so-called \emph{Carleman condition}
\begin{equation}
  \label{Carleman_Condition}
  \sum_{k=0}^{\infty} \, \mu_{k}^{-1/(2 k)} \; = \; \infty \, ,
\end{equation}
then the Pad\'{e} approximants $[m+j/m]_{f} (z)$ with $j \ge - 1$
constructed from the partial sums of the moment expansion
(\ref{Def_StieSer}) converge as $m \to \infty$ to the corresponding
Stieltjes function $f (z)$ \cite[Theorem
5.5.1]{Baker/Graves-Morris/1996}.

It can be shown that the Carleman condition (\ref{Carleman_Condition}) is
satisfied if the Stieltjes moments $\mu_{n}$ do not grow faster than
$C^{n + 1} (2 n)!$ as $n \to \infty$, where $C$ is a suitable positive
constant \cite[Theorem 1.3]{Simon/1982}. Thus, the Euler series
(\ref{Def:EulerSer}) satisfies the Carleman condition. An explicit proof,
that the Euler series satisfies the Carleman condition, was given in
\cite[pp.\ 239 - 240]{Baker/Graves-Morris/1996}.

Pad\'{e} approximants to a Stieltjes function $f (z)$ with $z > 0$ and $j
\ge -1$ satisfy several very useful inequalities \cite[Theorem
15.2]{Baker/1975}:
\begin{subequations}
  \label{PA_ineq_StieSer_1}  
  \begin{gather}
    \label{PA_ineq_StieSer_1_a}
    (-1)^{j+1} \, \bigl\{ [m+j+1/m+1]_{f} (z) - [m+j/m]_{f} (z) \bigr\} 
     \; \ge \; 0 \, ,
    \\
    \label{PA_ineq_StieSer_1_b}
    (-1)^{j+1} \, \bigl\{ [m+j/m]_{f} (z) - [m+j+1/m-1]_{f} (z) \bigr\} 
     \; \ge \; 0 \, ,
    \\
    \label{PA_ineq_StieSer_1_c}
    [m/m]_{f} (z) \ge f (z) \ge [m-1/m]_{f} (z) \, .
  \end{gather}
\end{subequations}
It follows from Eq.\ (\ref{PA_ineq_StieSer_1_a}) that for $z>0$ the
Pad\'{e} sequence $\bigl\{ [m+j/m]_{f} (z) \bigr\}_{m=0}^{\infty}$ is
\emph{increasing} if $j$ is \emph{odd}, and it is \emph{decreasing} if
$j$ is \emph{even}. In particular, the Pad\'{e} sequences $\bigl\{
[m+1/m]_{f} (z) \bigr\}_{m=0}^{\infty}$ and $\bigl\{ [m/m+1]_{f} (z)
\bigr\}_{m=0}^{\infty}$ are increasing, and the Pad\'{e} sequence
$\bigl\{ [m/m]_{f} (z) \bigr\}_{m=0}^{\infty}$ is decreasing.

If we set $j = - 1$ in Eq.\ (\ref{PA_ineq_StieSer_1_b}) and replace $m$
by $m+1$, we obtain for $z>0$ the inequality
\begin{equation}
  \label{PA_ineq_StieSer_2}
  [m/m+1]_{f} (z) \; \ge \; [m+1/m]_{f} (z) \, ,
  \qquad m \in \mathbb{N}_0 \, .
\end{equation}
Thus, the nesting sequences $\bigl\{ [m/m+1]_{f} (z)
\bigr\}_{m=0}^{\infty}$ and $\bigl\{ [m/m]_{f} (z)
\bigr\}_{m=0}^{\infty}$ provide the best lower and upper bounds to a
Stieltjes function $f (z)$ with positive argument. The fact that there
are nesting sequences of lower and upper bounds is of course extremely
helpful in actual computations since we immediately have an estimate of
the transformation error. These examples should suffice to show that
Pad\'{e} approximants possess a highly developed and practically useful
convergence theory in the case of Stieltjes series.

It is the purpose of this article to construct in the case of the Euler
series explicit expressions for the Pad\'{e} transformation error
$F^{[m/n]} (z)$ defined by
\begin{equation}
  \label{Pade_TransFormationError}
  [m/n]_{f} (z) \; = \; f (z) + F^{[m/n]} (z) \, ,
  \qquad m, n \in \mathbb{N}_{0} \, ,
\end{equation}
as well as analogous expressions for the delta transformation
(\ref{dWenTr}). Explicit expressions of the type of Eq.\
(\ref{Pade_TransFormationError}) are apparently extremely rare in the
theory of Pad\'{e} approximants (some counterexamples can be found in
articles by Allen, Chui, Madych, Narcowich, and Smith
\cite{Allen/Chui/Madych/Narcowich/Smith/1975}, Elliott
\cite{Elliott/1967}, Karlsson and von Sydow
\cite{Karlsson/vonSydow/1976}, and Luke \cite{Luke/1977b}).

To be practically useful, the transformation error $F^{[m/n]} (z)$ in
Eq.\ (\ref{Pade_TransFormationError}) should possess an explicit
expression of manageable complexity as a function of $m$, $n$, and
$z$. This is very important, because we usually have to construct simple
asymptotic approximations to $F^{[m/n]} (z)$ in order to understand how
the rate of convergence behaves if $m$ and $n$ become large. In this way,
we will not only be able to prove convergence of both Pad\'{e} and delta
in the case of the Euler series, but we will also obtain estimates for
the respective rates of convergence that hold in the limit of large
transformation orders.

The rarity of explicitly known Pad\'{e} transformation errors $F^{[m/n]}
(z)$ may well be a consequence of the fact that explicit expressions for
Pad\'{e} approximants are also very rare. In the books by Cuyt, Brevik
Petersen, Verdonk, Waadeland, and Jones
\cite{Cuyt/BrevikPetersen/Verdonk/Waadeland/Jones/2008}, Gil, Segura, and
Temme \cite{Gil/Segura/Temme/2007}, Luke \cite{Luke/1975}, and Sidi
\cite{Sidi/2003}, explicit expressions for the Pad\'{e} approximants of
some particularly convenient functions are listed. But in the vast
majority of all functions $f (z)$ possessing (formal) power series
expansions, neither the solutions to the systems of coupled linear
equations defining Pad\'{e} approximants \cite[Eqs.\ (1.5) and (1.7) on
pp.\ 2 and 3]{Baker/Graves-Morris/1996} nor their determinantal
expressions (\ref{PadeDetRep}) can be expressed in closed
form. Accordingly, Pad\'{e} approximants are -- as indicated by their
name -- normally \emph{numerical} expressions that belong to the realm of
approximation theory.

%
\typeout{==> Subsection: Wynn's Epsilon Algorithm}
\subsection{Wynn's Epsilon Algorithm}
\label{Sub:WynnEspAl}
%

It is generally agreed that the modern era of sequence transformations
started with two seminal articles by Shanks \cite{Shanks/1955} and Wynn
\cite{Wynn/1956a}, respectively. Shanks introduced in 1955 a powerful
sequence transformation that computes Pad\'e approximants. Wynn showed in
1956 that not only the Shanks transformation but also Pad\'{e}
approximants can be computed effectively by means of his celebrated
recursive epsilon algorithm. Of course, there is a very extensive
literature on these transformations. They are not only treated in books
on sequence transformations, but also in books on Pad\'{e}
approximants. In addition, there are countless articles describing
applications. These transformations are also treated in the recently
published NIST Handbook \cite[\S 3.9(iv) Shanks'
Transformation]{Olver/Lozier/Boisvert/Clark/2010}.

Wynn's epsilon algorithm corresponds to the following nonlinear recursive
scheme \cite[Eq.\ (4)]{Wynn/1956a} which is amazingly simple:
\begin{subequations}
  \label{eps_al}
  \begin{align}
    \label{eps_al_a}
    \varepsilon_{-1}^{(n)} & \; = \; 0 \, , 
    \qquad \varepsilon_0^{(n)} \, = \,
    s_n \, ,
    \qquad  n \in \mathbb{N}_{0} \, , \\
    \label{eps_al_b}
    \varepsilon_{k+1}^{(n)} & 
    \; = \; \varepsilon_{k-1}^{(n+1)} \, + \,
    \frac{1}{\varepsilon_{k}^{(n+1)} - \varepsilon_{k}^{(n)}} \, , 
    \qquad k, n \in \mathbb{N}_{0} \, .
  \end{align}
\end{subequations}
If the elements of the input sequence $\{ s_{n} \}_{n=0}^{\infty}$ are
the partial sums (\ref{ParSum_PowSer}) of a (formal) power series $f
(z)$, then the epsilon algorithm (\ref{eps_al}) produces Pad\'{e}
approximants to $f (z)$:
\begin{equation}
  \label{Eps_Pade}
  \varepsilon_{2 k}^{(n)} \; = \; [ k + n / k ]_f (z) \, ,
  \qquad k, n \in \mathbb{N}_{0} \, .
\end{equation}
It should be noted that Wynn's epsilon algorithm is not restricted to
input data that are the partial sums of a (formal) power series.
Therefore, it is actually more general and more widely applicable than
Pad\'{e} approximants. As a recent review we recommend
\cite{Graves-Morris/Roberts/Salam/2000a,*Graves-Morris/Roberts/Salam/2000b}.

If Wynn's epsilon algorithm is used to convert the partial sums of the
Stieltjes moment expansion (\ref{Def_StieSer}) to Pad\'{e} approximants
according to Eq.\ (\ref{Eps_Pade}), it makes sense to approximate the
corresponding Stieltjes function $f (z)$ by the following staircase
sequence in the Pad\'{e} table \cite[Eq.\ (4.3-7)]{Weniger/1989}:
\begin{align}
  \label{Pade_StairCaseSeq_EqsAlg}
  & \left\{ \varepsilon_{2\Ent{\nu/2}}^{\nu-2\Ent{\nu/2}}
  \right\}_{\nu=0}^{\infty} \; = \;
  \bigl\{ \bigl[ \Ent{(\nu+1)/2} / \Ent{\nu/2} \bigr] \bigr\}_{\nu=0}^{\infty}
  \notag
  \\
  & \qquad \; = \; [0/0], [1/0], [1/1], \cdots, [\nu/\nu], [\nu+1/\nu],
  [\nu+1/\nu+1], \cdots \, . 
\end{align}
Here, $\Ent{x}$ is the \emph{integral part} of the real number $x$, i.e.,
the largest integer $m$ satisfying $m \le x$.

It follows from Eqs.\ (\ref{PA_ineq_StieSer_1_c}) and
(\ref{PA_ineq_StieSer_2}) that the Pad\'{e} approximants in Eq.\
(\ref{Pade_StairCaseSeq_EqsAlg}) satisfy for $z>0$ the following
inequality \cite[Eq.\ (27)]{Bender/Weniger/2001}:
\begin{equation}
  \label{EpsPadeIneq}
  [m+1/m]_{f} (z) \; \le \; f (z) \; \le \; [m+1/m+1]_{f} (z) \,,
  \qquad m \in \mathbb{N}_{0} \, .
\end{equation}
Thus, the staircase sequence (\ref{Pade_StairCaseSeq_EqsAlg}) contains in
the case of a Stieltjes function $f (z)$ with $z>0$ two nesting sequences
$\varepsilon_{2m}^{(1)} = [m+1/m]_{f} (z)$ and $\varepsilon_{2m}^{(0)} =
[m/m]_{f} (z)$ of lower and upper bounds. Because of inequality
(\ref{PA_ineq_StieSer_2}), these bounds are less tight than those
provided by the nesting sequences $\bigl\{ [m/m+1]_{f} (z)
\bigr\}_{m=0}^{\infty}$ and $\bigl\{ [m/m]_{f} (z)
\bigr\}_{m=0}^{\infty}$ discussed in Section \ref{Sub:PadeApprox}.

With the help of some techniques developed in
\cite{Weniger/2000b,*Weniger/2000c}, inequality (\ref{EpsPadeIneq}) was
together with some other theoretical properties of Stieltjes series used
in \cite{Bender/Weniger/2001} to investigate numerically whether the
factorially divergent perturbation expansion for an energy eigenvalue of
the $\mathcal{P T}$-symmetric Hamiltonian $H (\lambda) = p^2 +
\nicefrac{1}{4} x^2 + \mathrm{i} \lambda x^3$ is a Stieltjes series,
which would imply its Pad\'{e} summability. Recently, the conjecture
formulated in \cite{Bender/Weniger/2001} was proven rigorously by
Grecchi, Maioli, and Martinez \cite{Grecchi/Maioli/Martinez/2009} (see
also \cite{Grecchi/Martinez/2013}).


%
\typeout{==> Section: Levin-Type Transformations}
\section{Levin-Type Transformations}
\label{Sec:Levin-TypeTransformations}
%

%
\typeout{==> Subsection: A General Construction Principle for Sequence
  Transformations}
\subsection{A General Construction Principle for Sequence
  Transformations}
\label{Sub:GenConstrPrincSeqTrans}
%

It is a typical feature of Wynn's epsilon algorithm (\ref{eps_al}) and of
many other sequence transformations that only the input of the numerical
values of a finite substring of a sequence $\{ s_{n} \}_{n=0}^{\infty}$
is required. No other information is needed to compute an approximation
to the (generalized) limit $s$ of the input sequence $\{ s_{n}
\}_{n=0}^{\infty}$. This is also true in the case of Pad\'{e}
approximants. For the computation of the Pad\'{e} approximant $[m/n]_{f}
(z)$, only the numerical values of the partial sums $f_{\nu} (z)$ with $0
\le \nu \le m+n$ are needed.

Very often, this it is a highly advantageous feature. However, in some
cases additional information on the index dependence of the truncation
errors $r_{n} = s_{n} - s$ is available. For example, the truncation
errors of Stieltjes series are according to Eqs.\
(\ref{Def_StieFunParSum}) and (\ref{StieSerErrEst}) bounded in magnitude
by the first term neglected in the partial sum, and they also have the
same sign patterns as the first terms neglected. It is obviously
desirable to utilize such a potentially very valuable \emph{structural}
information in order to enhance the efficiency of the transformation
process. Unfortunately, there is no obvious way of incorporating such an
information into Wynn's recursive epsilon algorithm (\ref{eps_al}). This
applies also to many other sequence transformations.

In 1973, Levin \cite{Levin/1973} introduced a sequence transformation
which overcame these limitations. It uses as input data not only a
sequence $\{ s_{n} \}_{n=0}^{\infty}$, which is to be transformed, but
also a sequence $\{ \omega_{n} \}_{n=0}^{\infty}$ of explicit
\emph{estimates} of the \emph{remainders} $r_{n} = s_{n} - s$. Levin's
transformation, which is also discussed in \cite[Chapter 3.9(v) Levin's
and Weniger's Transformations]{Olver/Lozier/Boisvert/Clark/2010}, is
generally considered to be one of the most powerful as well as most very
versatile sequence transformations currently known
\cite{Brezinski/RedivoZaglia/1991a,Sidi/2003,Smith/Ford/1979,%
  Smith/Ford/1982,Weniger/1989,Weniger/2004}. This explains why Levin's
$u$ transformation \cite[Eq.\ (7.3-5)]{Weniger/1989} is used internally
in the computer algebra system Maple to overcome convergence problems
(see for example \cite[pp.\ 51 and 125]{Corless/2002} or \cite[p.\
258]{Heck/2003}).

The derivation of Levin's sequence transformation becomes almost
trivially simple if we use a general construction principle for sequence
transformations introduced in \cite[Section 3]{Weniger/1989}. Here, we
present a slightly upgraded and formally improved version. Our starting
point is the model sequence
\begin{equation}
  \label{Mod_Seq_Om}
  s_{n} \; = \; 
  s \, + \, \omega_{n} z_{n}^{(k)} \, , \qquad k, n \in \mathbb{N}_0 \, .
\end{equation}
Obviously, this ansatz can only be useful if the products $\omega_{n}
z_{n}^{(k)}$ provide sufficiently accurate approximations to the
remainders $r_{n} = s_{n} - s$ of the sequence $\{ s_{n}
\}_{n=0}^{\infty}$ to be transformed.

The key quantities in Eq.\ (\ref{Mod_Seq_Om}) are the \emph{correction
  terms} $z_{n}^{(k)}$. The superscript $k$ characterizes their
complexity. In all examples considered here, $k$ corresponds to the
number of unspecified parameters occurring \emph{linearly} in
$z_{n}^{(k)}$. Since the remainder estimates $\omega_{n}$ are assumed to
be known, the approach based on Eq.\ (\ref{Mod_Seq_Om}) conceptually
boils down to the determination of the unspecified parameters in
$z_{n}^{(k)}$ and the subsequent elimination of $\omega_{n} z_{n}^{(k)}$
from $s_{n}$. Often, this approach leads to clearly better results than
the construction and elimination of other approximations to $r_{n}$.

As described in \cite[Section 3]{Weniger/1989}, there is a systematic
approach for the construction of a sequence transformation, which is
exact for the model sequence (\ref{Mod_Seq_Om}). Let us assume that a
class of \emph{linear} operators $\hat{T}_{k}$ can be found that
annihilate the correction terms $z_{n}^{(k)}$ according to $\hat{T}_{k}
\bigl( z_{n}^{(k)} \bigr) = 0$ for fixed $k$ and for all $n \in
\mathbb{N}_0$. We then obtain a sequence transformation, which is exact
for the model sequence (\ref{Mod_Seq_Om}), by applying $\hat{T}_{k}$ to
the ratio $[s_{n} - s] / \omega_{n} = z_{n}^{(k)}$. Since $\hat{T}_{k}$
annihilates $z_{n}^{(k)}$ and is by assumption also linear, the following
sequence transformation $\mathcal{T}_{k} (s_{n}, \omega_{n})$ \cite[Eq.\
(3.2-11)]{Weniger/1989} is exact for the model sequence
(\ref{Mod_Seq_Om}):
\begin{equation}
  \label{GenSeqTr}
  \mathcal{T}_{k} (s_{n}, \omega_{n}) \; = \; \frac
  {\hat{T}_{k} (s_{n} / \omega_{n} )} {\hat{T}_{k} (1 / \omega_{n} )} 
  \; = \; s \, .
\end{equation}
Originally, this approach was used for an almost trivially simple
derivation of Levin's transformation and for the construction of some
other, closely related sequence transformations \cite[Sections 7 -
9]{Weniger/1989}. Later, Brezinski and Redivo Zaglia
\cite{Brezinski/RedivoZaglia/1994a,Brezinski/RedivoZaglia/1994b} and
Brezinski and Matos \cite{Brezinski/Matos/1996} showed that this approach
is actually much more general than originally anticipated by Weniger, and
that the majority of the currently known sequence transformations can be
derived in this way. For further references on this topic, see \cite[p.\
1214]{Weniger/2004}.

As shown in \cite[Sections 7 - 9]{Weniger/1989} or in
\cite{Weniger/2004}, simple and yet very powerful sequence
transformations are obtained if the annihilation operator $\hat{T}_{k}$
in Eq.\ (\ref{GenSeqTr}) is based upon the finite difference operator
$\Delta$ defined by $\Delta f (n) = f (n+1) - f (n)$. Alternative
annihilation operators based on divided differences were discussed in
\cite[Section 7.4]{Weniger/1989}.

As is well known, $\Delta^{k}$ annihilates arbitrary polynomials
$\mathcal{P}_{k-1} (n)$ of degree $k - 1$ in $n$ according to $\Delta^{k}
\mathcal{P}_{k-1} (n) = 0$. Consequently, we choose $z_{n}^{(k)}$ in Eq.\
(\ref{Mod_Seq_Om}) in such a way that multiplication of $z_{n}^{(k)}$ by
some suitable quantity $W_k (n)$ yields a polynomial $\mathcal{P}_{k-1}
(n)$ of degree $k-1$ in $n$. If such pairs $z_{n}^{(k)}$ and $W_k (n)$
can be found, we obviously have
\begin{equation}
  \Delta^k \, [W_k (n) \, z_{n}^{(k)}] \; = \; 
  \Delta^k \, \mathcal{P}_{k-1} (n) \; = \; 0 \, .
\end{equation}
Thus, the weighted difference operator $\hat{T}_{k} = \Delta^k W_k (n)$
annihilates $z_{n}^{(k)}$.

As a further restriction, we consider here as in \cite[Section
II]{Weniger/2004} exclusively weights satisfying $W_k (n) = P_{k-1} (n)$,
where $P_{k-1} (n)$ is another polynomial of degree $k-1$ in $n$. Thus,
all Levin-type transformations considered here have the following general
structure:
\begin{equation}
  \label{GenLevTypeTr_DiffOpRep}
  T_{k}^{(n)} (s_{n}, \omega_{n}) \; = \; \frac 
  {\Delta^k \bigl\{ P_{k-1} (n) \, s_{n} / \omega_{n} \bigr\} } 
  {\Delta^k \bigl\{ P_{k-1} (n)/ \omega_{n} \bigr\} } \, ,
  \qquad k \in \mathbb{N} \, , \quad n \in \mathbb{N}_{0} \, .
\end{equation}
With the help of \cite[Eq.\ (25.1.1)]{Olver/Lozier/Boisvert/Clark/2010}
\begin{equation}
  \label{Delta^k_BinomSum}
  \Delta^k f(n) \; = \; (-1)^k \, \sum_{j=0}^k \, (-1)^j \, {\binom
    {k} {j}} \, f(n+j)\,,
\end{equation}
the right-hand side of Eq.\ (\ref{GenLevTypeTr_DiffOpRep}) can be
expressed as the ratio of two binomial sums:
\begin{equation}
  \label{GenLevTypeTr}
  T_{k}^{(n)} (s_{n}, \omega_{n}) \; = \; 
  \frac {\displaystyle \sum_{j=0}^{k} \, (-1)^{j} \,
    {\binom{k}{j}} \, \frac {P_{k-1} (n+j) \, s_{n+j}} {\omega_{n+j}}}
  {\displaystyle \sum_{j=0}^{k} \, (-1)^{j} \, \frac {P_{k-1} (n+j)}
    {\omega_{n+j}}} \, , \qquad k \in \mathbb{N} \, ,
  \quad n \in \mathbb{N}_{0} \, ,
\end{equation}
This is how Levin-type transformations are usually presented in the
literature (see for example \cite[Section II]{Weniger/2004}). But for our
convergence studies, the finite difference operator representation
(\ref{GenLevTypeTr_DiffOpRep}) is more useful than the explicit
expression (\ref{GenLevTypeTr}). Nevertheless, the ratio
(\ref{GenLevTypeTr}) has the undeniable advantage that we immediately
obtain an explicit expression for $T_{k}^{(n)} (s_{n}, \omega_{n})$ if
explicit expressions for the input sequence $\{s_{n} \}^\infty_{n=0}$ and
the remainder estimates $\{ \omega_{n} \}_{n=0}^{\infty}$ are known. For
most other transformations, explicit expressions are normally out of
reach.

In \cite[Section II]{Weniger/2004}, several different polynomials
$P_{k-1}$ were discussed which lead to different Levin-type
transformations. For these polynomials, the numerator and denominator
sums in Eq.\ (\ref{GenLevTypeTr}) can also be computed recursively
\cite[Section III]{Weniger/2004}. This is usually the best way of
numerically computing a Levin-type transformation.

The most important polynomials discussed in \cite[Section
II]{Weniger/2004} are the powers $P_{k-1} (n) = (\beta+n)^{k-1}$ with
$\beta > 0$, which yield Levin's sequence transformation
\cite{Levin/1973} in the notation of \cite[Eq.\ (7.1-7)]{Weniger/1989},
and the Pochhammer symbols $P_{k-1} (n) = (\beta+n)_{k-1} =
\Gamma(\beta+n+k-1)/\Gamma(\beta+n)$ with $\beta > 0$, which yield
Weniger's transformation \cite[Eq.\ (8.2-7)]{Weniger/1989}.

Let us assume that $s$ is the (generalized) limit of the sequence $\{
s_{n} \}_{n=0}^{\infty}$. Since the weighted difference operator
$\Delta^k P_{k-1} (n)$ is linear, the sequence transformation
(\ref{GenLevTypeTr_DiffOpRep}) can for all $k \in \mathbb{N}$ and for all
$n \in \mathbb{N}_{0}$ be expressed as follows:
\begin{equation}
  \label{GenLevTypeTr_DiffOpRepLimPlusRn}
  T_{k}^{(n)} (s_n, \omega_n) \; = \; s \, + \, \frac {\displaystyle
    \Delta^k \, \bigl\{ P_{k-1} (n) \, [s_{n} - s]/\omega_{n} \bigr\} } 
  {\displaystyle \Delta^k \, \bigl\{ P_{k-1} (n)/\omega_{n} \bigr\} }
  \; = \; s \, + \, \frac {\displaystyle 
    \Delta^k \, \bigl\{ P_{k-1} (n) \, r_{n}/\omega_{n} \bigr\} }
  {\displaystyle \Delta^k \, \bigl\{ P_{k-1} (n)/\omega_{n} \bigr\} } 
  \, . 
\end{equation}
This expression immediately tells how we can analyze the convergence of
convergence acceleration and summation processes. We have to investigate
whether and how fast the numerators and denominators of the ratio vanish
for fixed $n$ and as $k \to \infty$. Obviously, $T_{k}^{(n)} (s_n,
\omega_n)$ converges to $s$ if $\Delta^k P_{k-1} (n)$ annihilates
$[s_{n}-s]/\omega_{n} = r_{n}/\omega_{n}$ more effectively than
$1/\omega_{n}$.  Thus, we can expect good transformation results if
$[s_{n}-s]/\omega_{n}$ depends on $n$ less strongly than $1/\omega_n$.

%
\typeout{==> Subsection: Levin's and Weniger's Transformation}
\subsection{Levin's and Weniger's Transformation}
\label{Sub:LevinsAndWenigersTransformation}
%

Inverse power series play a prominent role in pure and applied
mathematics and also in the mathematical treatment of scientific and
engineering problems. Accordingly, it is an obvious idea to assume that
$z_{n}^{(k)}$ can be expressed as a \emph{truncated} inverse power series
in $\beta+n$:
\begin{equation}
  \label{CorrTerm_Levin}
  z_{n}^{(k)} \; = \; 
  \sum_{j=0}^{k-1} \, \frac{c_{j}^{(k)}}{(\beta+n)^{j}} \, , \qquad
  k \in \mathbb{N}, \quad n \in \mathbb{N}_{0} \, , \quad \beta > 0 \, . 
\end{equation}
Here and in analogous expressions occurring later it makes sense to
assume $c_{0}^{(k)} \ne 0$, but otherwise the unspecified coefficients
$c_{j}^{(k)}$ are in principle completely arbitrary.

The product $(\beta+n)^{k-1} \sum_{j=0}^{k-1} c_{j}^{(k)}/(\beta+n)^{j}$
is a polynomial of degree $k-1$ in $n$. Thus, it is annihilated by
$\Delta^{k}$, and we obtain in the notation of \cite[Eqs.\ (7.1-6) and
(7.1-7)]{Weniger/1989} the following expressions for Levin's sequence
transformation \cite{Levin/1973}:
\begin{align}
  \label{LevTrDiffOpRep}
  \mathcal{L}_{k}^{(n)} (\beta, s_{n}, \omega_{n}) & \; = \; 
    \frac {\Delta^{k} [(\beta+n)^{k-1} s_{n} / \omega_n]} 
     {\Delta^{k} [(\beta+n)^{k-1} / \omega_{n}]}
  \\
  \label{GenLevTr}
  & \; = \; \frac 
   {\displaystyle \sum_{j=0}^{k} \, (-1)^{j} \, {\binom{k}{j}} \, 
    \frac {(\beta+n+j)^{k-1}} {(\beta+n+k)^{k-1}} \, 
     \frac {s_{n+j}} {\omega_{n+j}}} 
   {\displaystyle \sum_{j=0}^{k} \, (-1)^{j} \, {\binom{k}{j}} \, 
    \frac {(\beta+n+j)^{k-1}} {(\beta+n+k)^{k-1}} \, 
     \frac {1} {\omega_{n+j}} } \, ,
   \qquad k, n \in \mathbb{N}_{0} \, , \quad \beta > 0 \, .
\end{align}

Different choices for $z_{n}^{(k)}$ lead to different sequence
transformations. We can for instance assume that $z_{n}^{(k)}$ can be
expressed as a truncated factorial series in $\beta+n$:
\begin{equation}
  \label{CorrTerm_Weniger}
  z_{n}^{(k)} \; = \; 
  \sum_{j=0}^{k-1} \, \frac{c_{j}^{(k)}}{(\beta+n)_{j}} \, , 
  \qquad k \in \mathbb{N}, \quad n \in \mathbb{N}_{0} \, , 
   \quad \beta > 0 \, . 
\end{equation}
Here, $(\beta+n)_{j} = \Gamma(\beta+n+j)/\Gamma(\beta+n)$ is a Pochhammer
symbol \cite[Eqs.\ (5.2.4) - (5.2.5)]{Olver/Lozier/Boisvert/Clark/2010}.

At first sight, the ansatz (\ref{CorrTerm_Weniger}) may look somewhat
strange. However, as briefly discussed in Appendix
\ref{App:FactorialSeries} or in more detail in \cite{Weniger/2010b},
there is a highly developed, but also largely forgotten theory of
factorial series. They play a similar role in the theory of difference
equations as inverse power series in the theory of differential
equations. Factorial series were fairly popular in the early twentieth
century, but now there is a deplorable lack of public awareness about
factorial series and their practical usefulness. Only a relatively small
group of specialists still uses them. However, we believe that this
widespread neglect of factorial series is not justified and that they can
be extremely useful mathematical tools.

Since the product $(\beta+n)_{k-1} \sum_{j=0}^{k-1}
c_{j}^{(k)}/(\beta+n)_{j}$ is a polynomial of degree $k-1$ in $n$, it is
annihilated by $\Delta^{k}$. We then obtain the following expressions for
the so-called $\mathcal{S}$ transformation \cite[Eqs.\ (8.2-6) and
(8.2-7)]{Weniger/1989}:
\begin{align}
  \label{WenTrDiffOpRep}
  \mathcal{S}_{k}^{(n)} (\beta, s_{n}, \omega_{n}) & \; = \; \frac
   {\Delta^{k} [(\beta+n)_{k-1} s_{n} / \omega_{n}]}  
   {\Delta^{k} [(\beta+n)_{k-1}/\omega_{n}]}
  \\
  \label{GenWenTr}
  & \; = \; \frac 
   {\displaystyle \sum_{j=0}^{k} \, (-1)^{j} \, {\binom{k}{j}} \, 
    \frac {(\beta+n+j)_{k-1}} {(\beta+n+k)_{k-1}} \, 
     \frac {s_{n+j}} {\omega_{n+j}} } 
   {\displaystyle \sum_{j=0}^{k} \, (-1)^{j} \, {\binom{k}{j}} \, 
    \frac {(\beta+n+j)_{k-1}} {(\beta+n+k)_{k-1}} \, 
     \frac {1} {\omega_{n+j}} } \, ,
   \qquad k, n \in \mathbb{N}_{0} \, , \quad \beta > 0 \, .
\end{align}
A highly condensed review of the historical development leading to the
derivation of this transformation was given in \cite[Section
2]{Weniger/2010b}). It is now common to call $\mathcal{S}_{k}^{(n)}
(\beta, s_n, \omega_n)$ and its variants the Weniger transformation (see
for example \cite{Borghi/2007,Borghi/2008c,Borghi/2009,Cvetic/Yu/2000,%
  Gil/Segura/Temme/2011,Li/Zang/Tian/2009,Temme/2007} or \cite[Eq.\
(9.53) on p.\ 287]{Gil/Segura/Temme/2007}). This terminology was also
used in the recently published NIST Handbook of Mathematical Functions
\cite[Chapter 3.9(v) Levin's and Weniger's
Transformations]{Olver/Lozier/Boisvert/Clark/2010}.

The transformation $\mathcal{S}_{k}^{(n)} (\beta, s_n, \omega_n)$ was
first used for the evaluation of auxiliary functions in molecular
electronic structure calculations \cite{Weniger/Steinborn/1989a}. Later,
predominantly the delta variant (\ref{Def:DeltaTasformation}), which will
be discussed later, was used with considerable success for the evaluation
of special functions and related objects
\cite{Jentschura/Gies/Valluri/Lamm/Weniger/2002,%
  Jentschura/Loetstedt/2012,Jentschura/Mohr/Soff/Weniger/1999,%
  Weniger/1989,Weniger/1990,Weniger/1992,Weniger/1994a,Weniger/1996d,%
  Weniger/2001,Weniger/2008,Weniger/Cizek/1990}, the summation of
divergent perturbation expansions \cite{Cizek/Vinette/Weniger/1991,
  Cizek/Vinette/Weniger/1993a,*Cizek/Vinette/Weniger/1993b,%
  Caliceti/Meyer-Hermann/Ribeca/Surzhykov/Jentschura/2007,%
  Cizek/Zamastil/Skala/2003,Jentschura/Becher/Weniger/Soff/2000,%
  Jentschura/Weniger/Soff/2000,Weniger/1990,Weniger/1992,Weniger/1994a,%
  Weniger/1996a,*Weniger/1996b,Weniger/1996c,Weniger/1996e,%
  Weniger/1997,Weniger/2004,Weniger/Cizek/Vinette/1991,%
  Weniger/Cizek/Vinette/1993}, and the prediction of unknown perturbation
series coefficients
\cite{Bender/Weniger/2001,Jentschura/Becher/Weniger/Soff/2000,%
  Jentschura/Weniger/Soff/2000,Weniger/1997}. More recently, the delta
transformation had also been employed in optics in the study of
nonparaxial free-space propagation of optical wavefields
\cite{Borghi/Santarsiero/2003,Li/Zang/Li/Tian/2009,Li/Zang/Tian/2009,%
  Dai/Li/Zang/Tian/2011,Borghi/Gori/Guattari/Santarsiero/2011} as well as
in the numerical evaluation of diffraction catastrophes
\cite{Borghi/2007,Borghi/2008a,Borghi/2008b,Borghi/2008c,%
  Borghi/2009,Borghi/2010a,Borghi/2011a,Borghi/2011b,Borghi/2012a}.

%
\typeout{==> Subsection: Drummond's Sequence Transformation}
\subsection{Drummond's Transformation}
\label{Sub:DrummondsSeqTr}
%

In the examples considered before, it was tacitly assumed that the
truncation errors $r_{n} = s-s_{n}$ of a slowly convergent or divergent
sequence can be approximated by the product $\omega_{n} z_{n}^{(k)}$, and
that the correction term $z_{n}^{(k)}$ approaches a constant as $n \to
\infty$. It is, however, possible to proceed differently.

The essential feature, which had been utilized in the previous examples,
is that polynomials of degree $k-1$ in $n$ are annihilated by
$\Delta^{k}$, Accordingly, we can also make the following ansatz:
\begin{equation}
  \label{CorrTerm_Drummond}
  z_{n}^{(k)} \; = \; 
  \sum_{j=0}^{k-1} \, c_j^{(k)} (\beta+n)^{j} \; = \; 
  \sum_{j=0}^{k-1} \, d_j^{(k)} n^{j} \, , \qquad
  k \in \mathbb{N}, \quad n \in \mathbb{N}_{0} \, , \quad \beta > 0 \, . 
\end{equation} 
Obviously, the corresponding annihilation operator $\hat{T}_{k}$ is
simply $\Delta^k$.

In \cite[Section 9.5]{Weniger/1989} it was shown that Drummond's sequence
transformation \cite{Drummond/1972} can in the notation of \cite[Eqs.\
(9.5-3) and (9.5-4)]{Weniger/1989} be derived via Eq.\
(\ref{CorrTerm_Drummond}), yielding
\begin{align}
  \label{DrummondTrDiffOpRep}
  \mathcal{D}_{k}^{(n)} (s_{n}, \omega_{n}) & \; = \; \frac 
   {\Delta^{k} \, [s_{n}/\omega_{n}]} {\Delta^{k} \, [1/\omega_{n}]}
  \\
  \label{Drummond_SeqTr}
  & \; = \; \frac 
   {\displaystyle \sum_{j=0}^{k} \, (-1)^{j} \, {\binom{k}{j}} \, 
    \frac {s_{n+j}} {\omega_{n+j}} } {\displaystyle 
     \sum_{j=0}^{k} \, (-1)^{j} \, {\binom{k}{j}} \, 
    \frac {1} {\omega_{n+j}} } \, ,
   \qquad k, n \in \mathbb{N}_{0} \, , \quad \beta > 0 \, .
\end{align}
Originally, Drummond had only considered the case $\omega_{n} = \Delta
s_{n}$ \cite[Eq.\ (1)]{Drummond/1972}. The more general transformation
$\mathcal{D}_{k}^{(n)} (s_{n}, \omega_{n})$ can be viewed to be a special
Levin-type transformation. Extensions of Drummond's sequence
transformation were recently discussed by Brezinski and Redivo Zaglia
\cite{Brezinski/RedivoZaglia/2010b}. 

Based on the model sequence $A_{r} = b_{r} T + c_{r} \sum_{i=0}^{n-1}
\gamma_{i} p_{i} (r)$ \cite[Eq.\ (1.16)]{Sidi/1981}, where $p_{i} (r)$
are polynomials of degree $i \le n-1$ in $r$, Sidi \cite[Eq.\
(1.17)]{Sidi/1981} derived a slightly more general transformation (see
also \cite[Lemma 17.3.2 on p.\ 327]{Sidi/2003}), and used it \cite[Eq.\ (1.17)]{Sidi/1981} for the
construction of an explicit expression for the Pad\'{e} approximant to
the divergent hypergeometric series ${}_{2} F_{0} (1, \mu; z)$ (see also
\cite[Section 13.3]{Weniger/1989}).


%
\typeout{==> Section: Convergence Analysis of the Delta Summation}
\section{Convergence Analysis of the Delta Summation}
\label{Sec:SummEulSerDeltaTrans}
%

%
\typeout{==> Subsection: Preliminaries}
\subsection{Preliminaries}
\label{Sub:SummEulSerDeltaTransPreliminaries}
%

Starting from the decomposition (\ref{ParSumEuFun=EuInt+TruncErr}), we
first have to choose the reaminder estimates $\{\omega_n\}^\infty_{n=0}$
of the Euler series. We recall that the remainder estimates
$\{\omega_{n}\}^\infty_{n=0}$ have to be chosen in such a way that
$\omega_{n}$ is proportional to the dominant term of the asymptotic
expansion of $r_{n}$ \cite[Eq.\ (7.3-1)]{Weniger/1989}, which implies here
\begin{equation}
  \label{Eq:AsymptoticsRemainder}
  \mathcal{R}_{n} (\mathcal{E}; z) \; = \; \omega_{n} \, 
  \bigl[ c + \mathrm{O} (1/n) \bigr] \, , \qquad n \to \infty \, .  
\end{equation}
The best simple estimate for the truncation error of a strictly
alternating convergent series is the first term not included in the
partial sum \cite[p.\ 259]{Knopp/1964}. Since the Euler series is a
Stieltjes series, its truncation error $\mathcal{R}_{n} (\mathcal{E}; z)$
is according to Eq.\ (\ref{EuSerTruncErrBound}) bounded in magnitude by
the first term neglected in the partial sum. Moreover, the first term
neglected is also an estimate of the truncation error of a divergent
hypergeometric series ${}_2 F_0 (a, b, - z)$ with $z > 0$, if its
parameters $a$ and $b$ satisfy $a+n, b+n > 0$ for some $n \in
\mathbb{N}_{0}$ \cite[Theorem 5.12-5]{Carlson/1977}. These examples all
suggest the remainder estimate
\begin{equation}
\label{dRemEst}
\omega_n \; = \; \Delta s_{n} \, ,
\end{equation}
which coincides with the one proposed by Smith and Ford \cite[Eq.\
(2.5)]{Smith/Ford/1979}. Using this remainder estimate in Eqs.\
(\ref{GenLevTr}) and (\ref{GenWenTr}) yields the following variants of
the sequence transformations $\mathcal{L}_{k}^{(n)} (\beta, s_n,
\omega_n)$ and $\mathcal{S}_{k}^{(n)} (\beta, s_n, \omega_n)$ \cite[Eqs.\
(7.3-9) and (8.4-4)]{Weniger/1989}:
\begin{align}
  \label{dLevTr}
  d_{k}^{(n)} (\beta, s_n) & \; = \; 
   \mathcal{L}_{k}^{(n)} (\beta, s_n, \Delta s_n) \, ,
  \\
  \label{dWenTr}
  \delta_{k}^{(n)} (\beta, s_n) & \; = \; 
   \mathcal{S}_{k}^{(n)} (\beta, s_n, \Delta s_n) \, .
\end{align}
Levin's $d$ transformation (\ref{dLevTr}) and the delta transformation
(\ref{dWenTr}) possess the following finite difference operator
representations \cite[Eqs.\ (7.1-6) and (8.2-6)]{Weniger/1989}:
\begin{align}
  \label{Def:dLevinTrDiffOpRep}
  \displaystyle
  d^{(n)}_{k} (\beta, s_{n}) & \; = \; \frac
  {\Delta^{k} \{(\beta+n)^{k-1} \, s_{n}/\Delta s_{n}\} }
  {\Delta^{k} \{(\beta+n)^{k-1}/ \Delta s_{n} \}} \, ,  
  \qquad k, n \in \mathbb{N}_{0} \, ,
  \\
  \label{Def:DeltaTasformation}
  \displaystyle
  \delta^{(n)}_k (\beta, s_{n}) & \; = \; \frac
  {\Delta^k\{(\beta+n)_{k-1} \, s_n/\Delta s_{n}\}}
  {\Delta^k\{(\beta+n)_{k-1}/ \Delta s_{n}\}} \, , 
  \qquad k, n \in \mathbb{N}_{0} \, .  
\end{align}
If the elements of the input sequence $\{ s_n \}_{n=0}^{\infty}$ are the
partial sums (\ref{ParSum_PowSer}) of a (formal) power series, then
$d_{k}^{(n)} \bigl( \beta, f_{n} (z) \bigr)$ and $\delta_{k}^{(n)} \bigl(
\beta, f_{n} (z) \bigr)$ can be expressed as ratios of two polynomials in
$z$ of degrees $k+n$ and $k$, respectively \cite[Eqs.\ (4.26) and
(4.27)]{Weniger/Cizek/Vinette/1993}:
\begin{align}
  \label{dLevTrPS}
  d_{k}^{(n)} \bigl(\beta, f_{n} (z) \bigr) & \; = \; \frac
  {\displaystyle
  \sum_{j=0}^{k} \, (-1)^{j} \, {\binom{k}{j}} \,
  \frac {(\beta + n +j )^{k-1}} {(\beta + n + k )^{k-1}} \,
  \frac {z^{k-j} f_{n+j} (z)} {\gamma_{n+j+1}} }
  {\displaystyle
  \sum_{j=0}^{k} \, (-1)^{j} \, {\binom{k}{j}} \,
  \frac {(\beta + n +j )^{k-1}} {(\beta + n + k )^{k-1}} \,
  \frac {z^{k-j}} {\gamma_{n+j+1}} } 
  \, , \qquad k, n \in \mathbb{N}_0 \, ,
  \\
  \label{dWenTrPS}
  \delta_{k}^{(n)} \bigl(\beta, f_{n} (z) \bigr) & \; = \; \frac
  {\displaystyle 
   \sum_{j=0}^{k} \, (-1)^{j} \, {\binom{k}{j}} \, 
   \frac {(\beta + n +j )_{k-1}} {(\beta + n + k)_{k-1}} \, 
   \frac {z^{k-j} f_{n+j} (z)} {\gamma_{n+j+1}} } 
  {\displaystyle 
   \sum_{j=0}^{k} \, (-1)^{j} \, {\binom{k}{j}} \, 
    \frac {(\beta + n + j)_{k-1}} {(\beta + n + k)_{k-1}} \, 
    \frac {z^{k-j}} {\gamma_{n+j+1}} } \, , 
   \qquad k, n \in \mathbb{N}_0 \,
\end{align}
The rational approximants $d_{k}^{(n)} \bigl(\beta, f_{n} (z) \bigr)$ and
$\delta_{k}^{(n)} \bigl(\beta, f_{n} (z) \bigr)$ satisfy the
accuracy-through-order relationships $f (z) - d_{k}^{(n)} \bigl(\beta,
f_{n} (z) \bigr) = \mathrm{O} \bigl( z^{k+n+2} \bigr)$ and $f (z) -
\delta_{k}^{(n)} \bigl(\beta, f_{n} (z) \bigr) \; = \; \mathrm{O} \bigl(
z^{k+n+2} \bigr)$ as $z \to 0$ \cite[Eqs.\ (4.28) and
(4.29)]{Weniger/Cizek/Vinette/1993}, which resemble the
accuracy-through-order relationship (\ref{Acc_Thr_Ord_1}) of Pad\'{e}
approximants. In \cite[Section 6]{Weniger/2004}, it was shown that
$d_{k}^{(n)} \bigl(\beta, f_{n} (z) \bigr)$ and $\delta_{k}^{(n)}
\bigl(\beta, f_{n} (z) \bigr)$ can also be viewed to be Pad\'{e}-type
approximants which are generalizations of Pad\'{e} approximants that were
introduced by Brezinski in his article \cite{Brezinski/1979} and fully
developed in his book \cite{Brezinski/1980a}.

As previously mentioned, the positive scaling parameter $\beta$ occurring
in $\mathcal{L}_{k}^{(n)} (\beta, s_{n}, \omega_{n})$ and
$\mathcal{S}_{k}^{(n)} (\beta, s_{n}, \omega_{n})$ and their variants is
customarily set to 1 in most practical applications. Thus, in the
following text we shall consider exclusively $\beta=1$.

It is the aim of our article to show that the delta transformation
applied to the partial sums of the Euler series converges for fixed $n
\in \mathbb{N}_{0}$ and for $z \in \mathbb{C} \setminus (-\infty, 0]$ to
the Euler integral:
\begin{equation}
  \label{OurTask}
  \mathcal{E} (z) \; = \;
  \lim_{k \to \infty} \, 
   \delta_{k}^{(n)} \bigl( 1, \mathcal{E}_{n} (z) \bigr) 
  \; = \; \lim_{k \to \infty} \, \frac
   {\displaystyle 
    \sum_{j=0}^{k} \, {\binom{k}{j}} \, 
     \frac {(n + j + 1)_{k-1}} {(n+j+1)!} \, 
      \sum_{\nu=0}^{n+j} \, (-1)^{\nu} \, \nu! \,z^{k-j+\nu} } 
   {\displaystyle 
    \sum_{j=0}^{k} \, {\binom{k}{j}} \, 
     \frac {(n + j + 1)_{k-1}}  {(n+j+1)!} \, z^{k-j} } \, .
\end{equation}

Our main technical problem is that both the numerators and denominators
of the delta transformation (\ref{Def:DeltaTasformation}) are binomial
sums of the type of $\Delta^kF_n = \sum_{j=0}^k (-1)^{k+j} {\binom{k}{j}}
F_{n+j}$. Unfortunately, it is in general extremely difficult to estimate
the asymptotics of such a binomial sum $\Delta^{k} F_{n}$ with fixed $n$
as $k\to\infty$ \cite{Flajolet/Sedgewick/1995,Flajolet/Sedgewick/2009},
let alone to find an explicit expression for $\Delta^{k} F_{n}$.

These technical problems can be overcome with the help of Borghi's
factorial series representation for the remainder of the Euler
series \cite[Eq.\ (52)]{Borghi/2010b}:
\begin{equation}
  \label{Eq:BorghiANM}
  \frac
  {\mathcal{R}_{n} (\mathcal{E}; z)}{(-1)^{n+1} (n+1)! z^{n+1}} 
  \; = \; - \,
  \sum_{k=0}^{\infty} \, \frac{L_{k}^{(-1)} (1/z)}{z} \, 
  \frac{k!}{(n+1)_{k+1}} \, .
\end{equation}
Here, $L_{n}^{(\alpha)} (x)$ is a generalized Laguerre polynomial
\cite[Eq.\ (18.5.12)]{Olver/Lozier/Boisvert/Clark/2010}. This expansion,
which extends and perfects the purely symbolic approach of \cite[Section
7]{Weniger/2007a}, is our central mathematical tool. With the help of the
factorial series (\ref{Eq:BorghiANM}), we will eventually arrive at a
closed form expression for the transformation error
\begin{equation}
  \label{Eq:DenominatorDelta.8}  
  \delta^{(n)}_{k} (1, \mathcal{E}_{n}(z)) - \mathcal{E}(z) \; = \;
  \frac 
   {\displaystyle
    \Delta^{k} \left\{ \frac
     {(n+1)_{k-1} \mathcal{R}_{n} (\mathcal{E}; z)}
      {(-1)^{n+1} (n+1)! z^{n+1}} \right\}}
   {\displaystyle \Delta^{k} \left\{ \frac
    {(n+1)_{k-1}}{(-1)^{n+1} (n+1)! z^{n+1}} \right\}} \, ,
\end{equation}
whose asymptotics for fixed $n$ as $k \to \infty$ can be analyzed.

%
\typeout{==> Subsection: Mimicking the Cut}
\subsection{Mimicking the Cut}
\label{Sub:AnalysisDenominator}
%

As shown in Eq.\ (\ref{dWenTrPS}), the delta transformation
$\delta_{k}^{(n)} \bigl(1, \mathcal{E}_{n}(z) \bigr)$ of the Euler series
is the ratio of two polynomials of degrees $k+n$ and $k$ in $z$,
respectively. Thus, it is a meromorphic function which is, with the
exception of the zeros of the denominator polynomial, analytic for all $z
\in \mathbb{C}$. In contrast, the Euler integral is analytic for all $z
\in \mathbb{C} \setminus (-\infty, 0]$, i.e., it has a cut along the
negative real semi-axis. Obviously, a rational function like
$\delta_{k}^{(n)} \bigl(1, \mathcal{E}_{n} (z) \bigr)$ cannot have such a
cut. But we can only hope for good approximations to the Euler integral,
if delta is somehow able to mimic the cut $(-\infty, 0]$. This can be
accomplished if all zeros and poles of $\delta_{k}^{(n)} \bigl(1,
\mathcal{E}_{n} (z) \bigr)$ are located in $(-\infty, 0]$.

The confinement of all the poles of $\delta_{k}^{(n)} \bigl(1,
\mathcal{E}_{n} (z) \bigr)$ to the negative real semi-axis $(-\infty, 0]$
is also important in connection with so-called \emph{spurious poles} (see
for example \cite{Stahl/1998} and references therein). Let us assume that
a given function with a finite number of singularities is to be
approximated by a sequence of rational functions such as continued
fractions, Pad\'{e}, or delta with increasing degrees of the numerator
and denominator polynomials. Then, the singularities of the function have
to be approximated somehow by the poles of the rational
function. However, not all poles of the rational functions necessarily
correspond to the singularities of the function which is to be
approximated. These spurious poles, which are essentially approximation
artifacts, can occur in regions where convergence is to be
expected. Obviously, spurious poles can easily frustrate any attempt of
formulating a rigorous convergence theory.

In the case of Pad\'{e} approximants $[M+J/M]$ with $J \ge -1$ to a
Stieltjes series, it is known that all their poles are simple, that they
lie on the negative real semi-axis, and that they have positive residues
(see for example \cite[Theorem 5.2.1 on p.\
201]{Baker/Graves-Morris/1996}). In this way, a Pad\'{e} approximant
simulates the cut of a Stieltjes function on the negative real
semi-axis. Therefore, we have to prove in the case of delta that
\emph{all} zeros of the denominator polynomials on the r.h.s. of Eq.\
(\ref{Eq:DenominatorDelta.8}) are \emph{real} and \emph{negative}.

It is comparatively easy to show that the denominator in Eq.\
(\ref{Eq:DenominatorDelta.8}) can be expressed as a terminating
hypergeometric series ${}_{2} F_{2}$ \cite[Eq.\
(16.2.1)]{Olver/Lozier/Boisvert/Clark/2010}. We only have to use Eq.\
(\ref{Delta^k_BinomSum}) together with ${\binom {k} {j}} =
(-1)^j\,{(-k)_j}/{j!}$ to obtain
\begin{equation}
  \label{App:DenominatorDelta.5}
  \Delta^{k} \left\{ \frac
    {(n+1)_{k-1}}{(-1)^{n+1} (n+1)! z^{n+1}} \right\} \; = \; 
  \frac {(-1)^{k+n+1}}{z^{n+1}} \, \frac{(n+1)_{k-1}}{(n+1)!} \, 
   {}_{2} F_{2}\left( \genfrac{}{}{0pt}{}{-k, k+n} {n+1, n+2};
    - \frac{1}{ z} \right) \, .
\end{equation}
We now have to prove that all zeros of this terminating ${}_{2} F_{2}$
are real and negative. Since all coefficients of this ${}_{2} F_{2}$ are
positive, it suffices to prove that all zeros are real.

This aim can be accomplished by considering some special generalized
hypergeometric series that are called \emph{P\'{o}lya frequency
  functions} \cite{Richards/1990,Schoenberg/1951}. In \cite[Theorem
4.1]{Richards/1990} it was shown that the generalized hypergeometric
series
\begin{equation}
  {}_{p} F_{q} \left( \genfrac{}{}{0pt}{}
   {\alpha_{1}+k_{1}, \dots, \alpha_{p}+k_{p}} 
    {\alpha_{1}, \dots, \alpha_{q}}; x \right)
\end{equation}
with $p \le q$, $\alpha_{1}, \dots, \alpha_{q} > 0$, and $k_{1}, \dots,
k_{p} \in \mathbb{N}$ is a P\'{o}lya frequency function, which according
to \cite[Lemma 5]{Driver/Jordaan/Martinez-Finkelshtein/2007} implies that
the associated terminating generalized hypergeometric series
\begin{equation}
  {}_{p+1} F_{q} \left( \genfrac{}{}{0pt}{}
   {-m, \alpha_{1}+k_{1}, \dots, \alpha_{p}+k_{p}} 
    {\alpha_{1}, \dots, \alpha_{q}}; -x \right)  
\end{equation}
with $m \in \mathbb{N}$ has only real zeros. If we now choose $p=1$,
$q=2$, $m=k$, $\alpha_{1}=n+1$, $k_{1}=k-1$, $\alpha_{2}=n+2$, and
$x=1/z$, we see that for all $n \in \mathbb{N}_{0}$ and for all $k \ge 2$
the terminating ${}_{2} F_{2}$ in Eq.\ (\ref{App:DenominatorDelta.5}) has
only negative zeros. Direct inspection shows that this is also true for
$k=1$. Thus, the denominator in (\ref{Eq:DenominatorDelta.8}) simulates
for all $k \in \mathbb{N}$ and for all $n \in \mathbb{N}_{0}$ the cut
$(-\infty, 0]$ of the Euler integral. We can also rule out spurious poles
of delta in $\mathbb{C} \setminus (-\infty, 0]$.

%
\typeout{==> Subsection: Integral Representation for the Delta
  Transformation Error}
\subsection{Integral Representation for the Delta Transformation Error}
\label{Sub:AnalysisDeltaTransError}
%

Next, we analyze the numerator in Eq.\ (\ref{Eq:DenominatorDelta.8}).
Our starting point is the factorial series expansion
(\ref{Eq:BorghiANM}) for the remainder of the Euler series, which -- as
explained in Appendix \ref{App:FactorialSeries} -- leads to the following
integral representation of $\mathcal{R}_{n} (\mathcal{E}; z)$:
\begin{subequations}
  \label{IntRep_Eq52_1}
  \begin{align}
    \label{IntRep_Eq52_1_a}
    \frac {\mathcal{R}_{n} (\mathcal{E};z)}{(-1)^{n+1} (n+1)! z^{n+1}} 
     & \; = \; - \frac {1}{z} \, \int_{0}^{1} \, 
      t^{n} \, \varphi_{\mathcal{E}} (t) \, \mathrm{d} t \, , 
    \\
    \label{IntRep_Eq52_1_b}	
    \varphi_{\mathcal{E}} (t) & \; = \; \sum_{k=0}^{\infty} \,   
     L_{k}^{(-1)} (1/z) \, (1-t)^{k} \, .  
  \end{align}
\end{subequations}
Equation (\ref{IntRep_Eq52_1}) alone would not be such a great
achievement. However, an explicit expression for the infinite series in
Eq.\ (\ref{IntRep_Eq52_1_b}) can be obtained by using the generating
function of the generalized Laguerre polynomials (see for example
\cite[Eq.\ (18.12.13)]{Olver/Lozier/Boisvert/Clark/2010}) which yields
\begin{equation}
  \label{Converging}
   \frac{\mathcal{R}_{n} (\mathcal{E}; z)}{(-1)^{n+1} (n+1)! z^{n+1}} 
  \; = \; - \frac {1}{z} \, \int_{0}^{1} \, \exp 
   \left( - \frac {1-t}{zt} \right) \, t^{n} \, \mathrm{d} t \, .
\end{equation}
If we substitute Eq.\ (\ref{Converging}) into the numerator of Eq.\
(\ref{Eq:DenominatorDelta.8}) and use Eq.\ (\ref{Delta^k_BinomSum}), we
obtain
\begin{align}
  \label{App:NumeratorDelta.1}
  & \Delta^{k} \left\{ \frac {(n+1)_{k-1} \mathcal{R}_{n} (\mathcal{E}; z)}
    {(-1)^{n+1} (n+1)! z^{n+1}} \right\} \; = \; - \frac{1}{z} \, 
     \int_{0}^{1} \, \exp \left( - \frac{1-t}{zt} \right) \, \Delta^{k}
      \bigl\{ (n+1)_{k-1} \, t^{n} \bigr\} \, \mathrm{d} t
  \notag
  \\ 
  & \qquad \; = \; - \frac{(-1)^{k}}{z} \, \int_{0}^{1} \, 
     \exp \left( - \frac{1-t}{zt} \right) \, 
      \sum_{j=0}^{k} \, (-1)^{j} \, {\binom {k} {j}} \, (n+j+1)_{k-1} \, 
       t^{n+j} \, \mathrm{d} t \, .  
\end{align}
The finite sum in the integral can be expressed as a terminating
hypergeometric series ${}_2F_1$:
\begin{equation}
  \label{App:NumeratorDelta.2}
  \sum_{j=0}^{k} \, (-1)^{j} \, {\binom {k} {j}} \, (n+j+1)_{k-1} \,
   t^{n+j} \; = \; (n+1)_{k-1} \, t^{n} \, {}_{2} F_{1}\left(
    \genfrac{}{}{0pt}{}{-k,k+n} {n+1}; t \right) \, .
\end{equation}
Accordingly, Eq.\ (\ref{App:NumeratorDelta.1}) can be expressed as
follows: 
\begin{align}
  \label{Eq:BorghiANM.2}
  & \Delta^{k} \left\{ 
     \frac {(n+1)_{k-1} \mathcal{R}_{n} (\mathcal{E}; z)}
      {(-1)^{n+1} (n+1)! z^{n+1}} \right\}
  \notag
  \\
  & \qquad \; = \; \frac {(-1)^{k}}{z} \, (n+1)_{k-1} \,
   \int_{0}^{1} \, t^{n} \, \exp \left( - \frac {1-t}{zt} \right) \,
    {}_{2} F_{1} \left( \genfrac{}{}{0pt}{}{-k, k+n} {n+1}; t \right) 
     \, \mathrm{d} t \, . 
\end{align}
If we now combine Eqs.\ (\ref{App:DenominatorDelta.5}) and
(\ref{Eq:BorghiANM.2}), we obtain the following explicit expression for
the transformation error for the delta summation of the Euler series:
\begin{equation}
  \label{Eq:BorghiANM.2.1}
  \delta^{(n)}_{k} \bigl(1, \mathcal{E}_{n} (z)\bigr) \, - \, 
   \mathcal{E}(z) \; = \; -(-z)^n \, (n+1)! \, \frac
    {\displaystyle \int_{0}^{1} \, t^{n} \, 
     \exp \left( - \frac {1-t} {zt} \right) \,
      {}_{2} F_{1} \left( \genfrac{}{}{0pt}{}{-k, k+n} {n+1}; t \right) 
       \, \mathrm{d} t}
    {\displaystyle {}_{2} F_{2} \left(
     \genfrac{}{}{0pt}{}{-k, k+n}{n+1, n+2}; - \frac{1}{z} \right)} \, .  
\end{equation}
%

%
\typeout{==> Subsection: Asymptotic Estimate of the Delta Convergence
  Rate}
\subsection{Asymptotic Estimate of the Delta Convergence Rate}
\label{Sub:AsymptoticEstimateDeltaConvergenceRate}
%

To the best of our knowledge, the integral representation
(\ref{Eq:BorghiANM.2.1}) is new and we consider it to be a major
achievement. Unfortunately, it does not solve our basic problem of
proving that the delta transformation sums the Euler series to the Euler
integral according to Eq.\ (\ref{OurTask}). We were not able to find an
explicit expression for the integral in Eq.\
(\ref{Eq:BorghiANM.2.1}). Therefore, it is by no means obvious how the
right-hand side of Eq.\ (\ref{Eq:BorghiANM.2.1}) behaves if the
transformation order $k$ becomes large. As a remedy, we try to construct
asymptotic approximations to the numerators and denominators in Eq.\
(\ref{Eq:BorghiANM.2.1}) which hold for fixed $n$ as $k \to \infty$.

In order to simplify notation and to make the resulting mathematical
expressions more readable, we now set $n=0$. This does not sacrifice too
much generality since it corresponds to the most important choice in
practical applications. We then obtain for the integral representation
(\ref{Eq:BorghiANM.2.1}):
\begin{equation}
  \label{Eq:BorghiANM.2.1.1}
  \delta^{(0)}_{k} \bigl(1, \mathcal{E}_{0} (z)\bigr) \, - \, 
   \mathcal{E}(z) \; = \; - \frac 
  {\displaystyle \int_{0}^{1} \, 
   \exp \left( - \frac {1-t} {zt} \right) \,
    {}_{2} F_{1} \left( \genfrac{}{}{0pt}{}{-k, k} {1}; t \right) \, 
     \mathrm{d} t}
  {\displaystyle {}_{2} F_{2} \left(
    \genfrac{}{}{0pt}{}{-k, k}{1, 2}; -  \frac{1}{z} \right)} \, .  
\end{equation}
For technical reasons, we make the variable substitution $t=(1+x)/2$,
which yields:
\begin{equation}
  \label{Eq:BorghiANM.2.1.1.1}
  \delta^{(0)}_{k} \bigl(1, \mathcal{E}_{0} (z)\bigr) \, - \, 
   \mathcal{E}(z) \; = \; - \frac {1}{2} \,
    \exp \left(\frac {1}{z} \right) \,\frac 
     {\displaystyle \int_{-1}^{1} \, \exp \left(-\frac{2/z}{1+x} \right) 
      \, {}_{2} F_{1} \left( \genfrac{}{}{0pt}{}{-k, k} {1}; 
       \frac{1+x}2 \right) \, \mathrm{d} x}
     {\displaystyle {}_{2} F_{2} \left( \genfrac{}{}{0pt}{}{-k, k}{1, 2}; 
      - \frac{1}{z} \right)} \, .  
\end{equation}

In the following text, we will analyze the asymptotic behavior of the
transformation error $\delta^{(0)}_{k} \bigl(1, \mathcal{E}_{0} (z)\bigr)
- \mathcal{E}(z)$ as $k \to \infty$ by analyzing separately the
asymptotic behavior of the numerators and denominators in Eq.\
(\ref{Eq:BorghiANM.2.1.1.1}). We accomplish this with the help asymptotic
approximations to hypergeometric functions that were introduced by Fields
\cite{Fields/1965} and by Fields and Luke
\cite{Fields/Luke/1963a,Fields/Luke/1963b}, respectively, and later
discussed in a very detailed way in Luke's book \cite[Chapter
7.4]{Luke/1969a}

For an asymptotic analysis of the denominator in Eq.\
(\ref{Eq:BorghiANM.2.1.1.1}) as $k \to \infty$, we start from the
asymptotic expansion \cite[Eq.\ 7.4.5(6) on p.\ 263]{Luke/1969a}:
\begin{align}
  \label{Luke/1969a_Eq_7.4.5(6)}
   & {}_{p+2} F_{q} \left(
    \genfrac{}{}{0pt}{}{-n, n+\lambda, \alpha_{p}} 
     {\rho_{q}}; -z \right) \; \sim \; \sum_{t=1}^{p} \,
      \frac{(n+\lambda)_{-\alpha_{t}}}{(n+1)_{\alpha_{t}}} \,
       \mathcal{L}_{p+2, q}^{(\alpha_{t})} 
        (z \mathrm{e}^{\mathrm{i} \delta \pi})
  \notag
  \\
  & \qquad + \frac {(2\pi)^{(1-\beta)/2} \Gamma (\rho_{q})}{\beta^{1/2}
    \Gamma (\alpha_{p})} \, \bigl( N^{\beta} z \bigr)^{\nu} \,  
     \exp \left( N z^{1/\beta} \beta - (a z/3) - \Omega (-z)/
      (N z^{1/\beta}) + \mathrm{O} (N^{-2}) \right) \, ,
  \notag
  \\
  & \qquad \qquad \vert \arg (z) \vert \le \pi - \epsilon \, , 
   \quad \epsilon > 0 \, , \quad  \delta = +(-) \quad \text{if} \quad
    \arg (z) \le (>) \; 0 \, .
\end{align}
The asymptotic variable $N \to \infty$ is defined by $N^{\beta} =
n(n+\lambda)$ with $\beta = q-p+1$. The remaining unspecified quantities
occurring here are defined in \cite[Eqs.\ 7.4.5(4) and 7.4.5(5) on p.\
262]{Luke/1969a}. For $p=0$, the finite sum in Eq.\
(\ref{Luke/1969a_Eq_7.4.5(6)}) is an empty sum which vanishes. By setting
$q=2$, $n = k$, $\lambda = 0$, $\rho_{1} = 1$ and $\rho_{2} = 2$, we
obtain $\beta = 3$, $ N = k^{2/3}$, $a = 1$, and $\gamma = - 2/3$.

In the exponential factor in Eq.\ (\ref{Luke/1969a_Eq_7.4.5(6)}), we
discard all contributions that vanish at least like $\mathrm{O} \bigl(
1/N \bigr)$ as $N \to \infty$. We obtain after some simple algebra:
\begin{equation}
  \label{Eq:BorghiANM.4}
  {}_{2} F_{2} \left( \genfrac{}{}{0pt}{} {-k, k} {1, 2};
    - \frac{1}{z} \right) \; \sim \; 
   \frac{(z/k^{2})^{2/3}}{2\pi \sqrt{3}} \,
  \exp \left(\frac {3 k^{2/3}} {z^{1/3}} - \frac {1} {3z} \right) \, ,
  \qquad k \to \infty \, . 
\end{equation}

The integral in the numerator of Eq.\ (\ref{Eq:BorghiANM.2.1.1.1}) is
more challenging. We first assume $z>0$. This is necessary to justify the
validity of the intermediate steps of our derivation. As is well known,
manipulations based the stationary phase approach or related techniques,
which we will employ in Eqs.\ (\ref{generalCaseDelta}) -
(\ref{NumeratorDeltaApprox}), tend to be very restrictive with respect to
the ranges of arguments and parameters for which their validity can be
guaranteed. Fortunately, it is often possible to show via analytic
continuation that final results obtained in this way are less demanding
and are valid in larger domains (compare also our discussion of
Fej\'{e}r's formula (\ref{GLAG_Asy_Fejer}) in Section
\ref{Sub:AsymptoticEstimatePadeConvergenceRate}).

The asymptotic variable $k$ occurs only in the hypergeometric function
${}_2F_1$ in the integral. Thus, we employ the following asymptotic
approximation \cite[Eq.\ 7.4.2(8) on p.\ 250]{Luke/1969a} that holds as
$N \to \infty$:
\begin{align}
  \label{Luke/1969a_Eq_7.4.2(8)}
  & {}_{p+2} F_{p+1} \left(
   \genfrac{}{}{0pt}{}{-n, n+\lambda, \alpha_{p}}      
    {\rho_{p+1}}; z \right) \; \sim \; 
    \sum_{t=1}^{p} \, \frac 
     {(n+\lambda)_{-\alpha_{t}}}{(n+1)_{\alpha_{t}}} \,
      \mathcal{L}_{p+2, p+1}^{(\alpha_{t})} (z) \, + \, 
       \frac {\Gamma (\rho_{p+1}) \, N^{2\gamma}} 
        {\Gamma (\alpha_{p}) \, \Gamma(1/2)} \, 
         \frac {[\sin\theta/2]^{2\gamma}}
          {[\cos\theta/2]^{2\gamma+\lambda}}    
  \notag
  \\
  & \quad \times 
   \exp \left\{ \frac{\varphi_{2} (\theta) + a_{2}}{N^{2}} + 
    \mathrm{O} \left( \frac{1}{N^{4}} \right) \right\} \, 
     \cos \left\{ N \theta + \pi \gamma + \frac{\varphi_{1} (\theta)}{N} 
       \, + \, \frac{\varphi_{3} (\theta)} {N^{3}} \, + \, 
        \mathrm{O} \left( \frac{1}{N^{5}} \right) \right\} \, ,
  \notag
  \\
  & \qquad \vert \arg (z) \vert \le \pi - \epsilon \, , 
   \quad \vert \arg (1-z) \vert \le \pi - \epsilon \, , 
	\quad \epsilon \; > \; 0 \, .
\end{align}
The asymptotic variable $N \to \infty$ is defined by $N^{2} =
n(n+\lambda)$. In addition, we have $\cos (\theta) = 1-2z$ or
equivalently $z=\sin^{2} (\theta/2)$. The remaining unspecified
quantities are defined in \cite[Eq.\ 7.4.2(9) on pages\ 251 and\
252.]{Luke/1969a}. As in Eq.\ (\ref{Luke/1969a_Eq_7.4.5(6)}), the finite
sum is for $p=0$ an empty sum which vanishes.

If we set in Eq.\ (\ref{Luke/1969a_Eq_7.4.2(8)}) $p=0$, $\lambda=0$,
$n=k$, $\rho_{q+1}=\{ 1 \}$, $z=(1+x)/2$, and if we neglect all
contributions that vanish at least like $\mathrm{O} \bigl( 1/N \bigr)$ as
$N \to \infty$, we obtain:
\begin{equation}
  \label{Eq:BorghiANM.4.2}
  {}_{2} F_{1} 
   \left( \genfrac{}{}{0pt}{}{-k, k} {1}; \frac{1+x}2 \right) \; \sim \; 
    \frac {1}{\sqrt{\pi k}} \, 
     \left[ \tan \bigl( \theta/2 \bigr) \right]^{-1/2} \,
      \cos \bigl( k \theta - \pi/4 \bigr) \, , \qquad k \to \infty \, .
\end{equation}
Here $\cos (\theta) =-x$. If we now take  into account 
that $x = \pi - \arccos (\theta)$, we obtain:
\begin{equation}
  \label{Eq:BorghiANM.4.3}
  {}_{2} F_{1} 
  \left( \genfrac{}{}{0pt}{} {-k, k} {1}; \frac{1+x}2 \right) 
   \; \sim \; \frac {(-1)^k} {\sqrt{\pi\,k}} \,	
    \left( \frac{1-x}{1+x} \right)^{1/4} \,
     \cos \left( k \arccos (x) + \frac{\pi}{4} \right) \, , 
      \qquad k\to\infty \, . 
\end{equation}
For  technical reasons it is advantageous to express the cosine in terms
of exponentials:
\begin{equation}
  \label{Eq:BorghiANM.5}
  {}_{2} F_{1} \left( \genfrac{}{}{0pt}{}{-k, k} {1}; \frac{1+x}2 \right) \; \sim \; 
	\frac {(-1)^k} {\sqrt{\pi  k}}\,
   \Re \left[ \left( \frac {1-x} {1+x} \right)^{1/4} \,
    \exp \left( \mathrm{i} k \arccos (x) + 
     \frac {\mathrm{i} \pi} {4} \right) \right] \, , 
  \qquad k \to \infty \, ,
\end{equation}
If we insert this asymptotic approximation into the integral in
Eq. (\ref{Eq:BorghiANM.2.1.1.1}), we obtain:
\begin{subequations}
  \label{generalCaseDelta.2.1}
  \begin{align}
    \label{generalCaseDelta.2.1_a}
    & \int_{-1}^{1} \, \exp \left( - \frac {2/z} {x+1} \right) \,
     {}_{2} F_{1} 
      \left( \genfrac{}{}{0pt}{}{-k, k} {1}; \frac{1+x}2 \right)
       \, \mathrm{d} x
    \notag
    \\
    & \qquad \; \sim \;
    \frac {(-1)^k} {\sqrt{\pi k}} \, \Re 
     \Bigl( \exp \left( \mathrm{i} \pi/4 \right)
      \mathcal{J}_{k}  \bigr( 1/z \bigr) \Bigr) \, ,
    \qquad k \to \infty \, ,
    \\
    \label{generalCaseDelta.2.1_b}
    & \mathcal{J}_{k} (\alpha) \; = \; \int_{-1}^{1} \, 
     \exp \left( - \frac{2 \alpha}{x+1} \right) \,   
      \left( \frac {1-x} {1+x} \right)^{1/4} \, \exp \Bigl( \mathrm{i} k
       \arccos (x) \Bigr) \, \mathrm{d} x \, .
  \end{align}
\end{subequations}
To estimate the integral $\mathcal{J}_{k} (\alpha)$ as $k \to \infty$, we
first set $x=\cos (t)$, which yields
\begin{equation}
  \label{generalCaseDelta.4}
  \mathcal{J}_{k} (\alpha) \; = \; \, 
  \int_{0}^{\pi} \, \sin (t) \, \sqrt{\tan (t/2)} \, 
  \exp \left( \mathrm{i} k t - \frac{\alpha}{\cos^{2} (t/2)} \right) \,  
  \mathrm{d} t \, ,
\end{equation}
and, after setting $\tau = \tan (t/2)$,
\begin{equation}
  \label{generalCaseDelta.6}
  \mathcal{J}_{k} (\alpha) \; = \; \, 4 \exp(- \alpha) \, 
   \int_{0}^{\infty} \, \frac {\tau^{3/2}} {(1+\tau^{2})^{2}} \, 
    \exp \bigl(-\alpha \tau^2 \bigr) \,
     \exp \bigl (\mathrm{i} 2k \arctan (\tau) \bigr) \, 
      \mathrm{d} \tau \, .
\end{equation}
In this form, the large $k$ behavior of the integral can be estimated
with the help of the stationary phase approach (see for example
\cite[Chapter II.3]{Wong/1989}). For that purpose, we rewrite Eq.\
(\ref{generalCaseDelta.6}) as follows:
\begin{subequations}
  \label{generalCaseDelta} 
  \begin{align}
    \label{generalCaseDelta.7}
    \mathcal{J}_k(\alpha) & \; = \; 4 \exp(-\alpha) \, 
     \int_{0}^\infty \, g(\tau) \, \exp \Bigl( f (\tau) \Bigr) \, 
      \mathrm{d} \tau \, ,
    \\
    \label{generalCaseDelta.8}
    f(\tau) & \; = \; \frac {3}{2} \, \log (\tau) \, - \, 
     \alpha \tau^2 \, + \, \mathrm{i} 2 k \, \arctan (\tau) \, ,
    \\
    \label{generalCaseDelta.8.1}
    g (\tau) & \; = \; \frac {1} {(1+\tau^2)^{2}} \, .
  \end{align} 
\end{subequations}
The four saddles are solutions of the equation $f' (\tau)=0$:
\begin{equation}
  \label{generalCaseDelta.9}
  4 \alpha \tau^{4} \, + \, (4 \alpha - 3) \tau^{2} \, - \, 
   4 \mathrm{i} k \tau \, + \, 3 \; = \; 0 \, .
\end{equation}
If we now make in Eq.\ (\ref{generalCaseDelta.9}) the ansatz $\tau = A
k^{\gamma}$, we obtain the following equation:
\begin{equation}
  \label{generalCaseDelta.10}
  4 \alpha A^{4} k^{4\gamma} \, + \, (4 \alpha - 3) A^{2}
  k^{2\gamma} \, - \, 4A \mathrm{i} k^{\gamma+1} 
\, + \, 3 \; = \; 0 \, .
\end{equation}
In the asymptotic limit $k \to \infty$, we obtain the following
approximations to the roots $\tau_{0},\ldots,\tau_{3}$:
\begin{subequations}
  \begin{align}
    \label{generalCaseDelta.11}
    \tau_{m} & \; \simeq \; \left( \frac{k}{\alpha} \right)^{1/3} \, 
     \exp \left( \frac{\mathrm{i}\pi} 6 + 
      \frac{\mathrm{i}2\pi {m}}3 \right) \, , 
       \quad m = 0, \dots, 2 \, ,
  \\
    \label{generalCaseDelta.12}
    \tau_{3} & \; \simeq \; \frac{3\mathrm{i}}{4k} \, .
  \end{align}
\end{subequations}
Extensive numerical tests showed that it is sufficient to include in Eq.\
(\ref{generalCaseDelta.6}) only the contribution from $\tau_0$. Since we
have
\begin{subequations}
 \begin{align}
   \label{generalCaseDelta.13}
   f_{0} & \; = \; f (\tau_{0}) \; = \; \frac{3}{2} \, 
    \log \left( (\mathrm{i}k/\alpha)^{1/3} \right) \, - \, \alpha
    (\mathrm{i}k/\alpha)^{2/3} \, + \, 
     \mathrm{i} 2k \arctan \left( (\mathrm{i}k/\alpha)^{1/3} \right)
   \notag
   \\
   & \; \sim \; \mathrm{i} \bigl[ k+1/4 \bigr] \pi \, + \, 
    \frac{1}{2} \log \bigl( k/\alpha \bigr) \, + \, \frac{2}{3} \alpha
     \, - \, 3 k^{2/3} (-\alpha)^{1/3} \, , \qquad k \to \infty \, ,
   \\
   \label{generalCaseDelta.13.1}
    f''_{0} & \; = \; f'' (\tau_{0}) \; = \; - \frac{2}{3 \tau_{0}^2{}} 
     \, - \, 2 \alpha \, - \, \frac{4 \mathrm{i} k \tau_{0}}
      {[1+\tau_{0}]^{2}} \; \sim \; - 6 \alpha \, , 
       \qquad k \to \infty \, ,
    \\
    \label{generalCaseDelta.13.1.1}
    g_{0} & \; = \; g (\tau_{0}) \; \sim \; 
     \bigl( \mathrm{i}k/\alpha \bigr)^{-4/3} \, ,
      \qquad k \to \infty \, ,
 \end{align}
\end{subequations}
we obtain after some long but in principle straightforward algebra:
\begin{align}
  \setlength{\jot}{6pt}
  \label{NumeratorDeltaApprox}
  & \int_{0}^{1} \, t^{n} \, 
   \exp \left( - \frac {1-t} {zt} \right) \,
    {}_{2} F_{1} \left( \genfrac{}{}{0pt}{}{-k, k+n} {n+1}; t \right) 
     \, \mathrm{d} t
  \notag
  \\[1ex]%
  & \qquad \sim \; 4 (-1)^k \, \bigl( 2/k \bigr)^{1/2} \, \exp(-\alpha)
   \, \mathrm{Re} \left\{ \exp \bigl( \mathrm{i}\pi/4 \bigr) 
    \, g_0 \, \exp(f_0)\bigl[ -f''_{0} \bigr]^{-1/2} \right\} 
  \notag
  \\[1ex]%
  & \qquad \sim \, \frac {4} {3^{1/2} \, z^{1/3} \, k^{4/3}}
   \, \exp \left(-\frac {1}{3z} - \frac{3 \, k^{2/3}}{2 \, z^{1/3}}
    \right) \, \cos \left( \frac{3^{3/2} \, k^{2/3}}{2 \, z^{1/3}} +
     \frac{\pi}{6} \right) \, , \qquad k \to \infty \, . 
\end{align}
As remarked above, this result was derived under the assumption $z >
0$. We nevertheless expect this approximation to work for complex $z \in
\mathbb{C} \setminus \{0\}$ thanks to analytic continuation. As shown in
Section \ref{Sec:NumericalExamples}, numerical tests confirm our
conjecture. 

Combining Eq.\ (\ref{NumeratorDeltaApprox}) with Eqs.\
(\ref{Eq:BorghiANM.2.1.1.1}) and (\ref{Eq:BorghiANM.4}) eventually
yields:
\begin{equation}
  \label{Eq:BorghiANM.12}
  \delta^{(0)}_{k} \bigl(1, \mathcal{E}_{0} (z) \bigr) \, - \, 
   \mathcal{E} (z) \; \sim \;
   \frac{4\pi}{z} \, \exp \left( \frac {1} {z} \right) \,  
    \exp \left(- \frac {9 \, k^{2/3}} {2 \, z^{1/3}} \right) \, 
     \cos \left( \frac {3^{3/2} \, k^{2/3}} {2 \, z^{1/3}} + 
      \frac{\pi}{6} \right) \, , \quad k \to \infty \, .  
\end{equation}
This is our main result. It not only proves for $n=0$ that the delta
transformation sums the factorially divergent Euler series to the Euler
integral according to Eq.\ (\ref{OurTask}), but it also provides an
asymptotic estimate of the rate of convergence as a function of the
transformation order $k$. To the best of our knowledge, the
transformation error estimate (\ref{Eq:BorghiANM.12}) is also the first
transformation error estimate for \emph{any} Levin-type transformation
and \emph{any} non-trivial input sequence that holds in the limit of
infinite transformation orders $k$.


%
\typeout{==> Section Convergence Analysis for Pad\'{e} Approximants}
\section{Convergence Analysis for Pad\'{e} Approximants}
\label{Sec:AnalysisTruncationErrorPade}
%

%
\typeout{==> Subsection: An Explicit Expression for the Pad\'{e}
  Transformation Error}
\subsection{An Explicit Expression for the Pad\'{e} Transformation Error}
\label{Sub:ExplExprPadeTransErr}
%

As reviewed in Section \ref{Sub:PadeApprox}, Pad\'{e} approximants to
Stieltjes series possess a highly developed convergence theory. The Euler
series is a Stieltjes series, and the Euler integral is the corresponding
Stieltjes function. Therefore, Pad\'{e} approximants sum for fixed $n \ge
-1$ the factorially divergent Euler series to the Euler integral
\cite[Theorem 5.5.1]{Baker/Graves-Morris/1996}:
\begin{equation}
  \label{EuIntLimPade}
  \mathcal{E} (z) \; = \; 
   \lim_{k \to \infty} \, [k+n/k]_{\mathcal{E}} (z) \, ,
    \qquad \vert \arg (z) \vert < \pi \, .
\end{equation}
However, an explicit expression for the transformation error
$[k+n/k]_{\mathcal{E}} (z) - \mathcal{E} (z)$ of the type of Eq.\
(\ref{Eq:BorghiANM.2.1}) seems to be unknown. We also want to derive
suitable asymptotic approximations to $[k+n/k]_{\mathcal{E}} (z) -
\mathcal{E} (z)$ which make it possible to compare quantitatively the
respective rates of convergence of Pad\'{e} and delta.

Sidi \cite[Eq.\ (2.18)]{Sidi/1981} (compare also \cite[Chapter
17.3]{Sidi/2003}) could show that the Pad\'{e} approximants
$[k+n/k]_{\mathcal{E}} (z)$ with $k, n \in \mathbb{N}_{0}$ of the Euler
series can be expressed by Drummond's sequence transformation
$\mathcal{D}_{k}^{(n)} (s_{n}, \omega_{n})$ \cite{Drummond/1972} with
remainder estimates $\omega_{n} = \Delta s_{n}$. This Levin-type
transformation possesses according to Eq.\ (\ref{DrummondTrDiffOpRep}) a
finite difference operator representation, which is more convenient for
our purposes, and according to Eq.\ (\ref{Drummond_SeqTr}) also a closed
form expression as the ratio of finite sums. Thus, the Pad\'{e}
approximants $[k+n/k]_{\mathcal{E}}(z)$ to the Euler series can be
expressed in terms of Drummond's sequence transformation as follows (see
also \cite[Eq.\ (13.3-5)]{Weniger/1989}):
\begin{align}
  \label{Pade.0}
  [k+n/k]_{\mathcal{E}}(z) & \; = \; 
  \mathcal{D}_{k}^{(n)} \bigl(\mathcal{E}_n, \Delta \mathcal{E}_n\bigr) 
  \; = \; \frac 
   {\Delta^{k} \bigl\{\mathcal{E}_{n} (z)/ \Delta \mathcal{E}_{n} (z)
     \bigr\}} 
   {\Delta^{k} \bigl\{ 1/ \Delta \mathcal{E}_{n} (z) \bigr\}} 
  \; = \; \frac 
   {\displaystyle \Delta^{k} \left\{\displaystyle \sum_{\nu=0}^{n}
    \frac{(-1)^{n+\nu+1} \nu! z^{\nu-n-1}}{(n+1)!} \right\}} 
   {\displaystyle \Delta^{k} \left\{ \frac 
    {1} {(-1)^{n+1} (n+1)! z^{n+1}} \right\}}
  \\
  \label{PadeEuSer<->Drummond} 
  & \; = \; \frac 
   {\displaystyle \sum_{j+0}^{k} \binom {k}{j} \frac {z^{k-j}}{(n+j+1)!}
    \sum_{\nu=0}^{n+j} (-1)^{\nu} \nu! z^{\nu}} 
   {\displaystyle \sum_{j+0}^{k} \binom {k}{j} \frac {z^{k-j}}{(n+j+1)!}} 
    \, , \qquad k, n \in \mathbb{N}_{0} \, . 
\end{align}

The fact, that the Pad\'{e} approximants to the Euler series can be
expressed by a certain Levin-type transformation, is very helpful for our
purposes. We can employ the same mathematical technology as in Section
\ref{Sec:SummEulSerDeltaTrans}.

Our first step is the following finite difference operator representation
for the transformation error which closely resembles Eq.\
(\ref{Eq:DenominatorDelta.8}):
\begin{equation}
  \label{Pade.1}
  [k+n/k]_{\mathcal{E}} (z) \, - \, \mathcal{E}(z) \; = \;
   \frac 
    {\displaystyle \Delta^{k} \left\{ 
     \frac {\mathcal{R}_{n} (\mathcal{E}; z)} 
      {(-1)^{n+1} (n+1)! z^{n+1}} \right\}}
    {\displaystyle \Delta^{k} \left\{ 
      \frac {1} {(-1)^{n+1} (n+1)! z^{n+1}} \right\}} \, .
\end{equation}

Next, we derive explicit expressions for the numerator and denominator on
the right-hand side of Eq.\ (\ref{Pade.1}). With the help of Eq.\ (5.5),
the denominator can be expressed as a terminating confluent
hypergeometric series ${}_{1} F_{1}$:
\begin{align}
  \label{DenominatorPade}
  {\displaystyle 
   \Delta^{k} \left\{   
    \frac {1} {(-1)^{n+1} (n+1)! z^{n+1}} \right\}} & \; = \;
   \frac {(-1)^k} {(-z)^{n+1} \, (n+1)!} \,
    \sum_{j=0}^{k} \, \frac {(-k)_{j}} {(n+2)_{j}} \, 
     \frac {(-z)^{-j}}{j!}
  \notag
  \\
  & \; = \; \frac {(-1)^k} {(-z)^{n+1} \, (n+1)!} \,
   {}_{1} F_{1} \bigl( -k; n+2; - 1/z \bigr) \, .
\end{align}
We do not have to analyze whether all zeros of the ${}_{1} F_{1}$ are
real and negative, since all poles of a Pad\'{e} approximant to a
Stieltjes series are simple and lie on the negative real semi-axis
\cite[Theorem 5.2.1]{Baker/Graves-Morris/1996}. But we could deduce this
directly from the fact that the ${}_{1} F_{1}$ in Eq.\
(\ref{DenominatorPade}) can be expressed as a generalized Laguerre
polynomial \cite[Eq.\ (18.5.12)]{Olver/Lozier/Boisvert/Clark/2010}:
\begin{equation}
  \label{Pade_1F1<->GLag}
   {}_{1} F_{1} \bigl( -k; n+2; - 1/z \bigr) \; = \;  
    \frac{k!}{(n+2)_{k}} \, L_{k}^{(n+1)} (-1/z) \; = \;  
     \frac{(n+1)!}{(k+1)_{n+1}} \, L_{k}^{(n+1)} \bigl( -1/z \bigr) \, .
\end{equation}
As is well known, all zeros of a generalized Laguerre polynomial
$L_{n}^{(\alpha)} (x)$ with $\alpha > - 1$ are real and positive (see for
example \cite[Theorem 4.5.1 on p.\ 116]{Beals/Wong/2010} or \cite[Chapter
10.17 on p.\ 204]{Erdelyi/Magnus/Oberhettinger/Tricomi/HTF2/1953b}).
Therefore, all zeros of $L_{k}^{(n+1)} (-1/z)$ must be real and negative.

If we now combine (\ref{DenominatorPade}) and (\ref{Pade_1F1<->GLag}), we
obtain for the denominator in Eq.\ (\ref{Pade.1}):
\begin{equation}
  \label{DenominatorPade_GLag}
  {\displaystyle 
   \Delta^{k} \left\{   
    \frac {1} {(-1)^{n+1} (n+1)! z^{n+1}} \right\}} \; = \; 
   \frac {(-1)^k} {(-z)^{n+1} \, (k+1)_{n+1}} \,
    L_{k}^{(n+1)} \bigl( -1/z \bigr) \, .  
\end{equation}

The derivation of an explicit expression for the numerator in Eq.\
(\ref{Pade.1}) is more demanding. We achieve some simplification by
reformulating the right-hand side of Eq.\ (\ref{Converging}) with the
help of the variable transformation $\xi=(1-t)/(zt)$:
\begin{equation}
  \label{Pade.2}
  \frac {\mathcal{R}_{n} (\mathcal{E}; z)} 
   {(-1)^{n+1} (n+1)! z^{n+1}} \; = \; - \int_{0}^{\infty} \,
    \frac {\exp(-\xi)} {(1 + z \xi)^{n+2}} \, \mathrm{d}\xi \, . 
\end{equation}
We now apply $\Delta$ under the integral sign and use Eq.\ (5.5). Then,
we obtain with the help of the binomial theorem \cite[Eq.\
(1.2.2)]{Olver/Lozier/Boisvert/Clark/2010}:
\begin{align}
  \label{Pade.3}
  &\Delta^{k} \, \left\{ \frac {\mathcal{R}_{n} (\mathcal{E}; z)} 
   {(-1)^{n+1} (n+1)! z^{n+1}} \right\} \; = \; - \int_{0}^{\infty} \,
    \frac {\exp(-\xi)} {(1 + z \xi)^{2}} \, \Delta^{k} \, \left\{ 
   \frac {1} {(1 + z \xi)^{n}} \right\} \, \mathrm{d}\xi
  \\
  & \qquad \; = \; (-1)^{k+1} \, \int_{0}^{\infty} \,
    \frac {\exp(-\xi)} {(1 + z \xi)^{n+2}} \, \sum_{j=0}^{k} \, (-1)^{j} 
     \binom {k} {j} \, \left\{ 
   \frac {1} {(1 + z \xi)^{j}} \right\} \, \mathrm{d}\xi 
  \\
  & \qquad \quad \; = \; (-1)^{k+1} \, \int_{0}^{\infty} \,
    \frac {\exp(-\xi)} {(1 + z \xi)^{n+2}} \, 
     \left\{ 1 - \frac {1} {1 + z \xi} \right\}^{k} \, \mathrm{d}\xi 
  \\
  & \qquad \qquad \; = \; - (-z)^{k} \, \int_{0}^{\infty} \,
    \frac {\xi^{k} \, \exp(-\xi)} {(1 + z \xi)^{k+n+2}} \, 
     \mathrm{d}\xi \, .
\end{align}
The last integral can be expressed in closed form in terms of a Kummer
$U$ function \cite[Eq.\ (13.4.4)]{Olver/Lozier/Boisvert/Clark/2010},
yielding:
\begin{equation}
  \label{Pade.4}
   \Delta^{k} \, \left\{ \frac {\mathcal{R}_{n} (\mathcal{E}; z)} 
    {(-1)^{n+1} (n+1)! z^{n+1}} \right\} \; = \; 
     (-1)^{k+1} \, \frac{k!}{z} \, U \bigl( k+1, -n, 1/z \bigr) \, .
\end{equation}
If we now substitute Eqs.\ (\ref{DenominatorPade_GLag}) and
(\ref{Pade.4}) into Eq.\ (\ref{Pade.1}), we obtain the following closed
form expression for the Pad\'{e} transformation error:
\begin{equation}
  \label{Pade.5}
  [k+n/k]_{\mathcal{E}} (z) \, - \, \mathcal{E}(z) \; = \; 
   (-z)^{n} \, (k+1)_{n+1} \, k! \, \frac 
    {U \bigl( k+1, -n, 1/z \bigr)}
     {L_{k}^{(n+1)} \bigl( -1/z \bigr)} \, .
\end{equation}
With the help of $U (a, -n, z) = z^{n+1} U (a+n+1, n+2, z)$ \cite[Eq.\
(13.2.11)]{Olver/Lozier/Boisvert/Clark/2010}, the Kummer function in Eq.\
(\ref{Pade.5}) can be expressed by $z^{-n-1} U (k+n+2, n+2, 1/z)$, which
yields
\begin{equation}
  \label{Pade.5_1}
  [k+n/k]_{\mathcal{E}} (z) \, - \, \mathcal{E}(z) \; = \; (-1)^{n} \,
   \frac{(k+1)_{n+1} \, k!}{z} \, \frac 
    {U \bigl( k+n+2, n+2, 1/z \bigr)}
     {L_{k}^{(n+1)} \bigl( -1/z \bigr)} \, .
\end{equation}
To the best of our knowledge, the explicit expressions (\ref{Pade.5}) and
(\ref{Pade.5_1}) for the Pad\'e transformation error of the Euler series
are new.

%
\typeout{==> Subsection: Asymptotic Estimate of the Pad\'{e} Convergence
  Rate}
\subsection{Asymptotic Estimate of the Pad\'{e} Convergence Rate}
\label{Sub:AsymptoticEstimatePadeConvergenceRate}
%

We want to proceed as we did it in the case of the delta transformation
in Section \ref{Sub:AsymptoticEstimateDeltaConvergenceRate}, i.e., we
want to determine independently the leading order asymptotics of the
numerators and denominators on the right-hand side of Eqs.\
(\ref{Pade.5}) or (\ref{Pade.5_1}) for fixed $n$ as $k \to \infty$.

For an analysis of the denominators in Eqs.\ (\ref{Pade.5}) and
(\ref{Pade.5_1}), we can use the following leading order asymptotic
approximation to a generalized Laguerre polynomial, which is commonly
called \emph{Fej\'{e}r's formula} (see for example \cite[Theorem 8.22.1
on p.\ 198]{Szegoe/1975}):
\begin{equation}
  \label{GLAG_Asy_Fejer}
  L_{n}^{(\alpha)} (x) \; = \; \frac 
   {\mathrm{e}^{\nicefrac{x}{2}} n^{\nicefrac{\alpha}{2} -
     \nicefrac{1}{4}}} 
    {\pi^{\nicefrac{1}{2}} x^{\nicefrac{\alpha}{2} +
     \nicefrac{1}{4}}} \, \left[ \cos \left( 2 \sqrt{n x} - 
      \frac{\alpha \pi}{2} - \frac{\pi}{4} \right) 
       + \mathrm{O} \left( n^{-\nicefrac{1}{2}} \right) \right] \, ,
       \qquad n \to \infty \, , \quad x > 0 \, .
\end{equation}
As remarked above, all zeros of a generalized Laguerre polynomial
$L_{n}^{(\alpha)} (x)$ with $\alpha > -1$ are located on the positive
real semi-axis (see for example \cite[Theorem 4.5.1 on p.\
116]{Beals/Wong/2010} or \cite[Chapter 10.17 on p.\
204]{Erdelyi/Magnus/Oberhettinger/Tricomi/HTF2/1953b}). The side
condition $x > 0$ implies that Fej\'{e}r's formula (\ref{GLAG_Asy_Fejer})
is valid in the oscillatory region $[0, \infty)$ of $L_{n}^{(\alpha)}
(x)$. However, for $x \in \mathbb{C} \setminus [0, \infty)$
$L_{n}^{(\alpha)} (x)$ is a monotonic function without zeros.

In the numerators of Eqs.\ (\ref{Pade.5}) and (\ref{Pade.5_1}), there
occurs a generalized Laguerre polynomial $L_{k}^{(n+1)} \bigl( -1/z
\bigr)$ with $z \in \mathbb{C} \setminus (-\infty, 0]$. This raises the
question whether Fej\'{e}r's formula (\ref{GLAG_Asy_Fejer}) is really
suited for our purposes if the argument of $L_{k}^{(n+1)} \bigl( -1/z
\bigr)$ does not belong to the oscillatory region $(-\infty, 0]$. The
behavior of $L_{k}^{(n+1)} \bigl( -1/z \bigr)$ in its oscillatory region
$-\infty < -1/z < 0$ is important because it enables the Pad\'{e}
approximant $[k+n/k]_{\mathcal{E}} (z)$ to mimic the cut of the Euler
integral, but for our convergence analysis it is of no interest.

With respect to these questions, the mathematical literature is
contradictory. Beals and Wong \cite[p.\ 116]{Beals/Wong/2010} emphasized
that the convergence of Fej\'{e}r's formula (\ref{GLAG_Asy_Fejer}) is
uniform in $x$ for any compact interval $0 < \delta \le x \le
\delta^{-1}$, but they did not say anything about its behavior away from
the positive real semi-axis. Similarly, Erd\'{e}lyi, Magnus,
Oberhettinger, and Tricomi \cite[p.\
199]{Erdelyi/Magnus/Oberhettinger/Tricomi/HTF2/1953b}, Ismail \cite[p.\
118]{Ismail/2005}, and Szeg\"{o} \cite[p.\ 198]{Szegoe/1975} remarked
that Fej\'{e}r's formula (\ref{GLAG_Asy_Fejer}) is uniformly valid in any
compact subset of $[0, \infty)$. In contrast, Buchholz \cite[p.\
137]{Buchholz/1969} stated that Fej\'{e}r's formula
(\ref{GLAG_Asy_Fejer}) is valid for $0 \le \arg (x) < 2\pi$.

This apparent inconsistency can be resolved if we replace in Fej\'{e}r's
formula (\ref{GLAG_Asy_Fejer}) the cosine by exponentials via $\cos
(\varphi) = [\exp (\mathrm{i} \varphi) + \exp (-\mathrm{i}
\varphi)]/2$. This yields:
\begin{align}
  \label{GLag_Tricomi_AsyHankel_1}
  & L_{n}^{(\alpha)} (x) \; = \; 
   \frac {\mathrm{e}^{\nicefrac{x}{2}} n^{\nicefrac{\alpha}{2} -
     \nicefrac{1}{4}}} {2\pi^{\nicefrac{1}{2}} x^{\nicefrac{\alpha}{2} +
      \nicefrac{1}{4}}}
  \notag
  \\  
  & \qquad \times \, 
   \left[ \exp \left( \mathrm{i} \left\{ 2\sqrt{n x} - 
    \frac{\pi \alpha}{2} - \frac{\pi}{4} \right\} \right) + 
     \exp \left( - \mathrm{i} \left\{ 2\sqrt{n x} - \frac{\pi \alpha}{2} 
      - \frac{\pi}{4} \right\} \right) +
     \mathrm{O} \left( \frac{1}{n^{1/2}} \right) \right] \, ,
     \notag
     \\   
     & \qquad \quad n \to \infty, 
      \quad \vert \arg (2\sqrt{n x}) \vert < \pi \, .  
\end{align}
For $x > 0$, this is simply Fej\'{e}r's formula (\ref{GLAG_Asy_Fejer}) in
disguise. But for $x < 0$, we get something new because now only the
second exponential in (\ref{GLag_Tricomi_AsyHankel_1}) matters. The first
one is exponentially subdominant as $n \to \infty$ and can be
neglected. For $x < 0$, the branches of of $(-x)^{-\nicefrac{\alpha}{2} -
  \nicefrac{1}{4}}$ and $(-x)^{1/2}$ have to chosen in such a way that
they are real and positive.

Thus, for an analysis of the large index asymptotics of $L_{k}^{(n+1)}
\bigl( -1/z \bigr)$, which occurs in the denominators in Eqs.\
(\ref{Pade.5}) and (\ref{Pade.5_1}), we use
(\ref{GLag_Tricomi_AsyHankel_1}) which provides the following leading
order approximation:
\begin{equation}
  \label{Pade.6.0}
  L_{k}^{(n+1)} \bigl( -1/z \bigr) \; \sim \; 
   \frac{(z/\pi)^{1/2}}{2} \, 
    \exp \left(-\frac {1}{2z} + \frac {2 k^{1/2}} {z^{1/2}} \right) \,
     (z k)^{(2n+1)/4} \, , \qquad k \to \infty \, .
\end{equation}
This is essentially a special case of \emph{Perron's formula}
\cite[Theorem 4.8.9 on p.\ 118]{Ismail/2005} (a more general expression
was given in \cite[Theorem 8.22.3 on p.\ 199]{Szegoe/1975}):
\begin{equation}
  \label{GLAG_Asy_Perron}
  L_{n}^{(\alpha)} (x) \; = \; \frac 
   {\mathrm{e}^{\nicefrac{x}{2}} n^{\nicefrac{\alpha}{2} -
     \nicefrac{1}{4}}} 
    {\pi^{\nicefrac{1}{2}} (-x)^{\nicefrac{\alpha}{2} +
     \nicefrac{1}{4}}} \, \exp \left( 2 \sqrt{- n x} \right) \, ,
      \qquad n \to \infty \, , 
       \quad x \in \mathbb{C} \setminus (0, \infty) \, .
\end{equation}
Consequently, the leading order asymptotic approximation
(\ref{GLag_Tricomi_AsyHankel_1}) interpolates between Fej\'{e}r's formula
(\ref{GLAG_Asy_Fejer}) and Perron's formula (\ref{GLAG_Asy_Perron}).

If we also use the asymptotic approximation $\Gamma(z+a)/\Gamma(z+b) =
z^{a-b} \bigl[ 1 + \mathrm{O} (1/z) \bigr]$ as $z \to \infty$ \citep[Eq.\
(5.11.12)]{Olver/Lozier/Boisvert/Clark/2010}, which yields $1/(k+1)_{n+1}
\sim k^{-(n+1)}$ as $k \to \infty$, we obtain for the denominators in
Eqs.\ (\ref{Pade.5}) and (\ref{Pade.5_1}):
\begin{equation}
  \label{Pade.6.0}
  \frac{L_{k}^{(n+1)} \bigl( -1/z \bigr)}{(k+1)_{n+1}} \; \sim \;
   \frac 1{2\sqrt\pi} \,
    \exp \left(-\frac 1{2z} + \frac{2k^{1/2}}{z^{1/2}} \right) \,
     \frac{z^{(2n+3)/4}}{k^{(2n+3)/4}} \, \qquad k \to \infty \, .
\end{equation}

Concerning the numerator in Eq.\ (\ref{Pade.5}), we start from the
following leading order asymptotic approximation \cite[Eq.\
(13.8.8)]{Olver/Lozier/Boisvert/Clark/2010}, 
\begin{subequations}
  \label{KummerU_Asy_a->inf}  
  \begin{align}
    \label{KummerU_Asy_a->inf_a}
    U (a, b, x) & \; \sim \; \frac{\mathrm{e}^{x/2}}{\Gamma (a)} \, 
     \left[ \sqrt {\frac{2 \tanh (w/2)}{\beta}} \, 
      \left( \frac{1 - \mathrm{e}^{-w}}{\beta} \right)^{-b} \,
       \beta^{1-b} \, K_{1-b} (2 \beta a) \right. 
    \notag
    \\ 
    & \qquad \qquad \; + \; \left. \frac{1}{a} \, \left( 
     \frac{a^{-1} + \beta}{1 + \beta} \right)^{1-b} \, 
      \mathrm{e}^{-2 \beta a} \, \mathrm{O} (1) \right] \, ,
       \qquad a \to \infty \, ,
  \\
  \label{KummerU_Asy_a->inf_b}
  w & \; = \; \mathrm{arccosh} \left( 1 + \frac{x}{2a} \right) \, ,
   \qquad \beta \; = \; \frac{ w + \sinh (w)}{2} \, .
  \end{align}
\end{subequations}
which holds uniformly with respect to $x$ in $[0, \infty)$ for fixed $b
\le 1$. Here, $K_{1-b} (2 \beta a)$ is a modified Bessel function of the
second kind \cite[Eq.\ (10.27.4)]{Olver/Lozier/Boisvert/Clark/2010}.

If we now set in Eq.\ (\ref{KummerU_Asy_a->inf}) $a = k+1$, $b = - n$,
and $x = 1/z$, we obtain with the help of \cite[Eqs.\ (4.33.1) and
(4.38.4)]{Olver/Lozier/Boisvert/Clark/2010} the following asymptotic
estimates:
\begin{subequations}
  \label{Pade.6.2}
  \begin{align}
  \label{Pade.6.2_a}    
    w & \; = \; \mathrm{arccosh} \left( 1 + \frac {1}{2kz} \right)
     \; \sim \; \frac {1}{\sqrt{zk}} \, , \qquad k \to \infty \, , 
  \\
  \label{Pade.6.2_b}
  \beta & \; = \; \frac{w + w \bigl[ 1 + \mathrm{O (w)} \bigr]}{2} 
   \; = \; w \bigl[ 1 + \mathrm{O (w)} \bigr] \, , 
    \qquad w \to 0 \, ,
  \notag
  \\
  & \; \sim \; \frac {1}{\sqrt{zk}} \, , \qquad k \to \infty \, .
  \end{align}
\end{subequations}
If we now insert the asymptotic estimates in Eq.\ (\ref{Pade.6.2}) into
Eq.\ (\ref{KummerU_Asy_a->inf}), we obtain:
\begin{equation}
  \label{Pade.6.3}
  k! \, U \bigl( k+1, -n, 1/z \bigr) \; \sim \; 2 \sqrt{\pi} \, 
   \exp \left( \frac {1}{2z} \right) \, (zk)^{-({n+1})/2} \,
    K_{n+1} \left( 2 \sqrt{k/z} \right) \, , \qquad k \to \infty \, ,
\end{equation}
If we now approximate the modified Bessel function $K_{n+1} \left( 2
  \sqrt{k/z} \right)$ by the leading term of its asymptotic expansion
(\ref{BesK_AsySer}), we obtain the following asymptotic estimate for the
numerator in Eq.\ (\ref{Pade.5}):
\begin{equation}
  \label{Pade.7}
  k! \, U \bigl( k+1, -n, 1/z \bigr) \; \sim \; \sqrt{\pi} \, 
   \exp \left( \frac{1}{2z} - \frac{2 k^{1/2}}{z^{1/2}}  \right) \,
    {z^{-(2n+1)/4}} \, {k^{-(2n+3)/4}} \, , \qquad k \to \infty \, .
\end{equation}
Finally, we insert Eqs.\ (\ref{Pade.6.0}) and (\ref{Pade.7}) into Eq.\
(\ref{Pade.5}) to obtain the following leading order asymptotic
approximation to the Pad\'{e} transformation error:
\begin{equation}
  \label{Pade.8}
  [k+n/k]_{\mathcal{E}} (z) \, - \, \mathcal{E}(z) \; \sim \; 
   (-1)^{n} \, \frac{2\pi}{z} \, \exp \left(\frac{1}{z} \right) \, 
    \exp \left(- \frac{4 \, k^{1/2}}{z^{1/2}} \right)  \, , 
     \quad k \to \infty \, .
\end{equation}
It is remarkable that the $n$-dependence of the leading order estimate
(\ref{Pade.8}) of the transformation error of the Pad\'e approximant
$[k+n/k]_{\mathcal{E}}(z)$ occurs only in the form of the sign factor
$(-1)^n$ which decides whether the Pad\'e sequence
$\{[k+n/k]\}^\infty_{k=0}$ provides upper or lower bounds. This is in
agreement with Stieltjes inequalities (\ref{PA_ineq_StieSer_1}) discussed
in Section \ref{Sub:PadeApprox}. Moreover, we can conclude that all
Pad\'{e} sequences $[k+n/k]_{\mathcal{E}} (z)$ with $n \ge 0$ converge
with essentially the same rate to the Euler integral.  The differences in
the convergence rates of the Pad\'e sequences $[k+n/k]_{\mathcal{E}}(z)$
and $[k+n'/k]_{\mathcal{E}}(z)$, respectively, occur only in subdominant
terms.

If we set in Eq.\ (\ref{Pade.8}) $n=0$, we obtain the following leading
order asymptotic approximation to the transformation error of diagonal
Pad\'{e} approximants:
\begin{equation}
  \label{Eq:BorghiANM.14}
  [k/k]_{\mathcal{E}} (z) \, - \, \mathcal{E}(z) \; \sim \; 
  \frac{2\pi}{z} \, \exp \left(\frac{1}{z} \right) \, 
  \exp \left(- \frac{4 \, k^{1/2}}{z^{1/2}} \right)  \, , 
  \quad k \to \infty \, .
\end{equation}
A comparison of our asymptotic estimates (\ref{Eq:BorghiANM.12}) and
(\ref{Eq:BorghiANM.14}) shows that the rate of convergence of a diagonal
Pad\'{e} approximant $[k/k]_{\mathcal{E}} (z)$, whose behavior is
characterized by the decay term $\exp (- 4 k^{1/2}/z^{1/2})$, is much
lower than that of the delta transformation $\delta^{(0)}_{k} \bigl(1,
\mathcal{E}_{0} (z) \bigr)$ characterized by $\exp (-9 k^{2/3}/[2
z^{1/3}])$. To the best of our knowledge, this result constitutes the
first {theoretical} explanation of the well known superiority of the
delta transformation over Pad\'{e} approximants with respect to the
summation of factorially divergent alternating series. In addition, Eqs.\
(\ref{Eq:BorghiANM.12}) and (\ref{Eq:BorghiANM.14}) also provide
theoretical estimates of the respective rates of convergence. This
information seems to be missing in the current literature.


%
\typeout{==> Section: Numerical Results}
\section{Numerical Results}
\label{Sec:NumericalExamples}
%

%
\typeout{==> Subsection: Transformation Errors for Positive Arguments}
\subsection{Transformation Errors for Positive Arguments}
\label{Sub:TransfErrPosArg}
%

The derivation of our asymptotic transformation error estimates
(\ref{Eq:BorghiANM.12}) and (\ref{Eq:BorghiANM.14}) required several
partly drastic approximations whose validity is only guaranteed in the
asymptotic domain of very large transformation orders $k$. Therefore, it
is by no means obvious whether our estimates produce meaningful results
also for only moderately large or even small transformation orders
$k$. We check this question numerically.

In our analytical manipulations in Sections
\ref{Sub:AsymptoticEstimateDeltaConvergenceRate} and
\ref{Sub:AsymptoticEstimatePadeConvergenceRate} we had always assumed $z
> 0$. But our subsequent numerical results will demonstrate that our
estimates also work for complex arguments $z$ as long as we are not too
close to the cut $(-\infty, 0]$ of the Euler integral.

\begin{figure}[!ht]
  \centering
  \includegraphics[width=10cm]{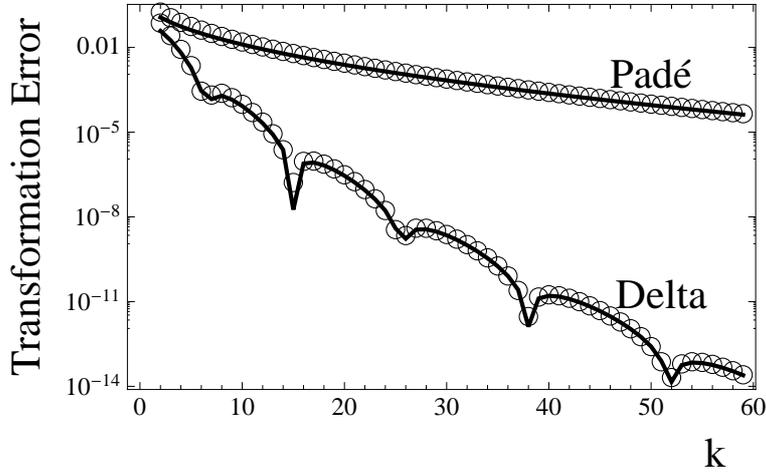}
  \caption{Comparison of the observed transformation errors (open
    circles) vs the transformation order $k$ of the sequences $\bigl\{
    \delta^{(0)}_k \bigr\}$ and $\bigl\{ [k/k] \bigr\}$ for the summation
    of the Euler series at $z=10$. The solid curves represent the corresponding
    asymptotic estimates (\ref{Eq:BorghiANM.12}) and
    (\ref{Eq:BorghiANM.14}), respectively.}
  \label{Fig:FigDeltaOnEulerz10}
\end{figure}

Figure \ref{Fig:FigDeltaOnEulerz10} displays the transformation errors
for Pad\'e and delta (open circles) vs the transformation order $k$ of
the sequences $\{\delta^{(0)}_k\}$ (labeled as ``delta'') and $[k/k]$
(labeled as ``Pad\'{e}''), for the summation of the Euler series at
$z=10$ (since the Euler series is asymptotic to the Euler integral as $z
\to 0$, an argument $z=10$ represents a very challenging summation
problem). The solid curves represent the corresponding asymptotic
estimates (\ref{Eq:BorghiANM.12}) and (\ref{Eq:BorghiANM.14}),
respectively.

The results in Fig.\ \ref{Fig:FigDeltaOnEulerz10} show that our
asymptotic estimates (\ref{Eq:BorghiANM.12}) and (\ref{Eq:BorghiANM.14})
also work quite well for \emph{small} transformation orders $k$, i.e.,
far away from the asymptotic regime $k \to \infty$. Several other
numerical experiments not shown here, which used different
\emph{positive} values of the argument $z$ of the ES, confirmed our
conclusion about the usefulness of the asymptotic estimates
(\ref{Eq:BorghiANM.12}) and (\ref{Eq:BorghiANM.14}) away from the
asymptotic regime.

%
\typeout{==> Subsection: Transformation Errors for Complex Arguments}
\subsection{Transformation Errors for Complex Arguments}
\label{Sub:TransfErrPosArg}
%

\begin{figure}[!ht]
  \centering
  \subfigure{
  \includegraphics[width=6cm]{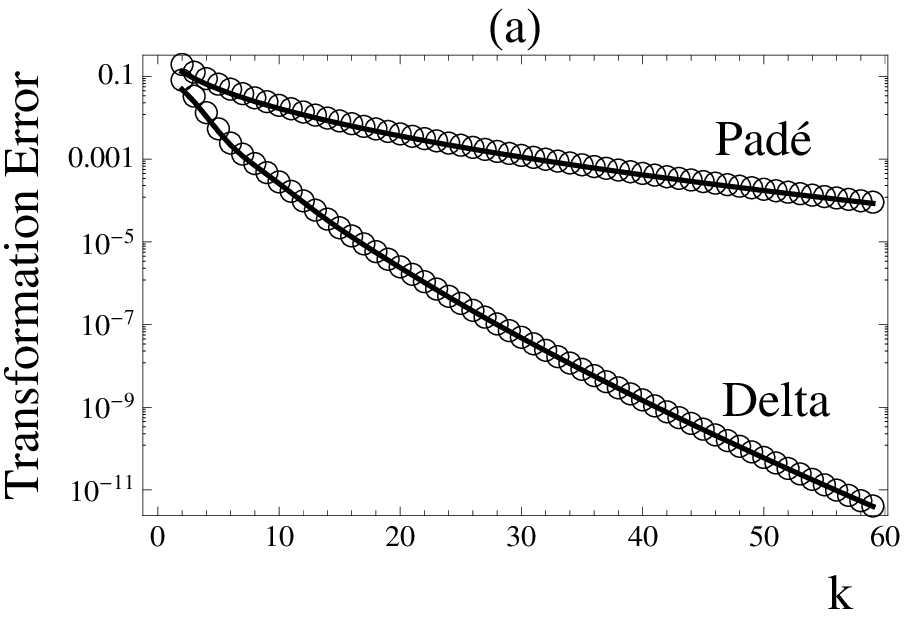}
  \hfill
  \includegraphics[width=6cm]{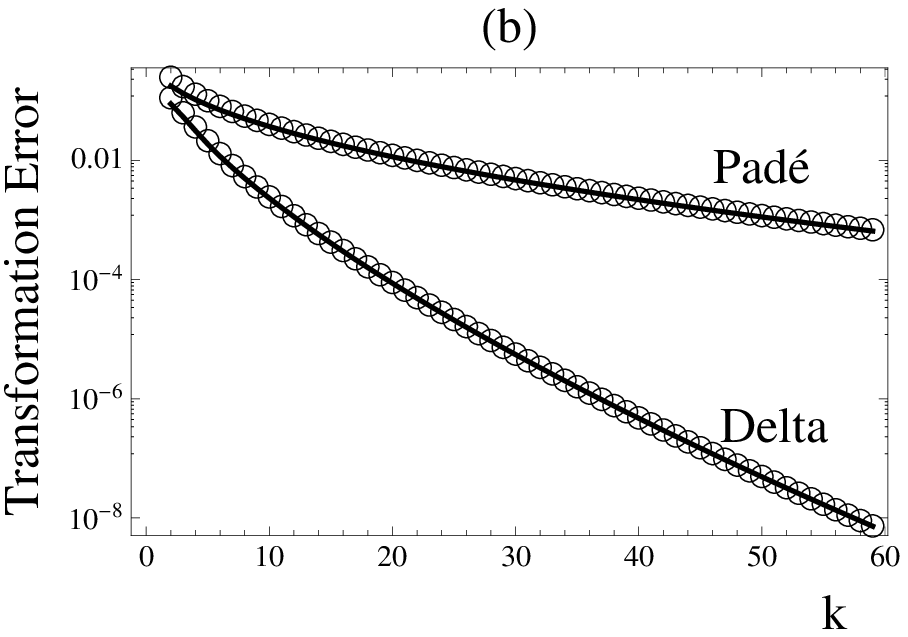}
  }
  \subfigure{
  \includegraphics[width=6cm]{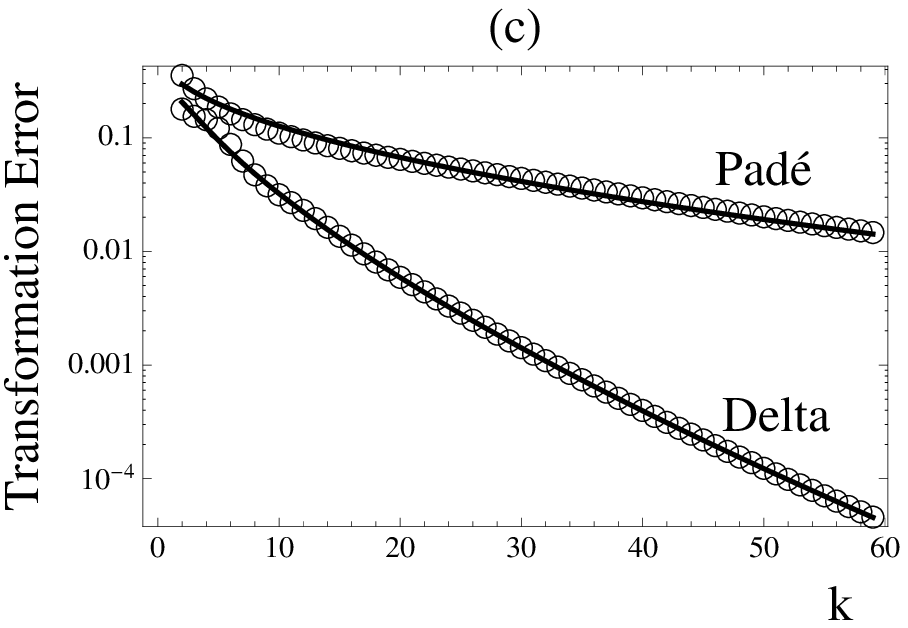}
  \hfill
  \includegraphics[width=6cm]{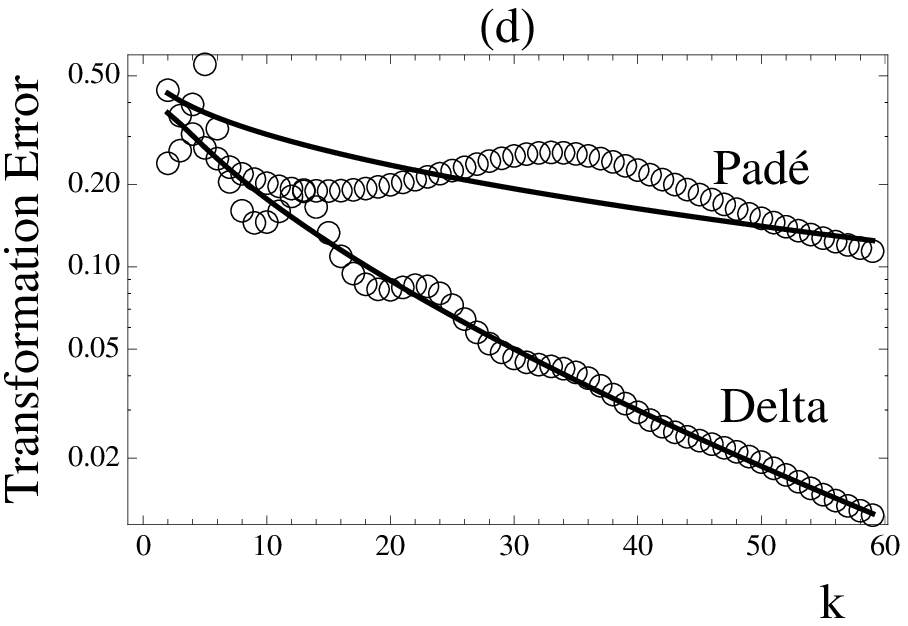}
  }
  \caption{The same as in Fig.\ \ref{Fig:FigDeltaOnEulerz10}, but for
    complex $z=10\,\exp(\mathrm{i}\varphi)$ with $\varphi=\pi/4$ (a),
    $\varphi=\pi/2$ (b), $\varphi=3\pi/4$ (c), and $\varphi=9\pi/10$
    (d).}
  \label{Fig:FigDeltaAndPadeOnEulerz10Complex}
\end{figure}

The plots in Fig.\ \ref{Fig:FigDeltaAndPadeOnEulerz10Complex} also
display the observed transformation errors vs $k$ as in Fig.\
\ref{Fig:FigDeltaOnEulerz10}, but now for \emph{complex} arguments $z =
\vert z\vert \exp (\mathrm{i} \varphi)$. The transformation
errors corresponding to $z = 10 \, \exp(\mathrm{i}\varphi)$ are plotted
for several values of $\varphi \in (0, \pi)$.

The plots in Fig.\ \ref{Fig:FigDeltaAndPadeOnEulerz10Complex} show that
our asymptotic estimates (\ref{Eq:BorghiANM.12}) and
(\ref{Eq:BorghiANM.14}) also work well for complex arguments $z$ of the
Euler series, provided that $z = \vert z \vert \exp (\mathrm{i} \varphi)$
does not get too close to the cut $(-\infty, 0]$ of the Euler integral,
which corresponds to $\varphi = \pm \pi$. This happens in Fig.\
\ref{Fig:FigDeltaAndPadeOnEulerz10Complex}.d. But in all plots, the
superiority of the delta transformation over Pad\'{e} approximants is
evident.

The rates of convergence displayed in Fig.\
\ref{Fig:FigDeltaAndPadeOnEulerz10Complex} assume their maxima for $z>0$
($\Longleftrightarrow \varphi = 0$) and decrease monotonically in
magnitude as $z = \vert z \vert \exp(\mathrm{i} \varphi)$ approaches the
cut $(-\infty, 0]$ ($\Longleftrightarrow \varphi = \pm \pi$). Both for
delta and for Pad\'{e}, these observations can be explained by analyzing
our asymptotic estimates (\ref{Eq:BorghiANM.12}) and
(\ref{Eq:BorghiANM.14}). We distinguish three cases (i): $z > 0
\Longleftrightarrow \varphi = 0$, (ii): $z \in \mathbb{C} \setminus
(-\infty, 0] \Longleftrightarrow \varphi \in (-\pi, \pi) \setminus \{ 0
\}$, and (iii): $z \in (-\infty, 0] \Longleftrightarrow \varphi = \pm
\pi$.

If $z > 0$, the convergence of delta is guaranteed by the exponential
factor in our asymptotic estimate (\ref{Eq:BorghiANM.12}). The cosine
term in Eq.\ (\ref{Eq:BorghiANM.12}) only produces some modulation, which
can be observed in Fig.\ \ref{Fig:FigDeltaOnEulerz10}, but which do not
affect the essentially exponential decay. Similarly, the convergence of
Pad\'{e} is for $z > 0$ guaranteed by our asymptotic estimate
(\ref{Eq:BorghiANM.14}).

If $z$ belongs to the cut complex plane $\mathbb{C} \setminus (-\infty,
0]$, the situation is much more complicated. Because of $\cos (x) =
\bigl[ \exp (\mathrm{i} x) + \exp (-\mathrm{i} x) \bigr]/2$, the cosine
factor in Eq.\ (\ref{Eq:BorghiANM.12}) can for $\varphi \in (-\pi, \pi)
\setminus \{ 0 \}$ no longer be ignored in our convergence analysis. If
we transform the cosine in Eq.\ (\ref{Eq:BorghiANM.12}) to complex
exponentials and multiply it by the exponential in Eq.\
(\ref{Eq:BorghiANM.12}) -- which is best done with the help of a computer
algebra system like Maple or Mathematica -- we obtain exponentially
decaying terms modulated by sine and cosine terms which, however,
guarantee the convergence of delta in the cut complex plane $\mathbb{C}
\setminus (-\infty, 0]$. Similarly, our asymptotic estimate
(\ref{Eq:BorghiANM.14}) implies that Pad\'{e} converges in the cut
complex plane $\mathbb{C} \setminus (-\infty, 0]$.

But if $z$ lies on the cut $(-\infty, 0]$, the situation is quite
different. If we again transform the cosine in Eq.\
(\ref{Eq:BorghiANM.12}) by complex exponentials and multiply it by the
exponential in Eq.\ (\ref{Eq:BorghiANM.12}), we now obtain for $\varphi =
\pm \pi$ in addition to exponentially decaying terms also sine and cosine
terms which are \emph{not} damped by exponentially decaying terms. Thus,
delta does not converge if $z$ lies on the cut $(-\infty, 0]$. Such a
divergence occurs also in the case of Pad\'{e}.

This negative result is actually no surprise. As discussed before,
rational approximants like delta or Pad\'{e} can only mimic the cut
$(-\infty, 0]$ of the Euler integral but not exactly reproduce it. In
addition, the Euler series is for $z \in (-\infty, 0]$ an extremely
demanding summation problem since the Euler integral is for $z \in
(-\infty, 0]$ a double-valued function. This follows at once from the
fact that the Euler integral is according to Eq.\
(\ref{Def:EulerIntegral}) essentially the exponential integral $E_1$ with
argument $1/z$, But for negative real argument the exponential integral
$E_1 (z)$ is also double-valued. It satisfies \cite[Eq.\
(6.5.1)]{Olver/Lozier/Boisvert/Clark/2010}:
\begin{equation}
  \label{E1Cut}
  E_1 (-z \pm \mathrm{i}0) \; = \;
  - \, \mathrm{Ei} (z) \, \mp \, \mathrm{i} \pi \, , \qquad z > 0 \, .
\end{equation}
Here, $\mathrm{Ei}$ is another exponential integral which for $z>0$ is
essentially a \emph{Cauchy principal value integral} \cite[Eqs.\ (6.2.5)
and (6.5.2)]{Olver/Lozier/Boisvert/Clark/2010}. If we replace in Eq.\
(\ref{E1Cut}) $z$ by $1/(-z \mp \mathrm{i}0) = -1/z \pm \mathrm{i}0$ with
$z > 0$ and combine the resulting expression with Eq.\
(\ref{Def:EulerIntegral}), we obtain \cite[Eqs.\ (28) -
(30)]{Weniger/2007d}:
\begin{align}
  \int_{0}^{\infty} \, \frac{\exp (-t) \mathrm{d}t} {1+(-z \mp \mathrm{i}0)t} & \; = \; 
	\frac
   {\exp \bigl(1/(-z \mp \mathrm{i}0) \bigr)} 
   {-z \mp \mathrm{i}0} \, E_1 \bigl(1/(-z \mp \mathrm{i}0) \bigr)
   \notag
   \\
   & \; = \; \frac {\exp(-1/z)} {z} \, \bigl[ \mathrm{Ei} (1/z)
    \, \mp \, \mathrm{i} \pi \bigr] \, , \qquad z > 0 \, .
\end{align}
This expression shows that $\mathcal{E} (-z \mp \mathrm{i}0)$ is
double-valued since it contains a nonzero imaginary part $\mp \pi
\exp(-1/z)/z$, whose sign depends on how we approach the negative real
axis and which is nonanalytic as $z \to 0$.

%
\typeout{==> Subsection: Summation of the Euler Series by Levin's Transformation}
\subsection{Summation of the Euler Series by Levin's Transformation}
\label{Sub:LevTr}

In view of the similarity of Levin's $\mathcal{L}$ and Weniger's
$\mathcal{S}$ transformation, a skeptical reader might wonder why we do
not also consider the $d$-variant of the Levin's 
transformation defined in Eq. (\ref{dLevTr}) whose finite 
difference operator representation is \cite[Eq.\ (7.3-9)]{Weniger/1989}
\begin{equation}
  \label{Def:dLevinTrDiffOpRep}
  d^{(n)}_{k} (\beta, s_{n}) \; = \; \frac
  {\Delta^{k} \{(\beta+n)^{k-1} \, s_{n}/\Delta s_{n}\} }
  {\Delta^{k} \{(\beta+n)^{k-1}/ \Delta s_{n} \}} \, , 
  \qquad k, n \in \mathbb{N}_{0} \, ,
\end{equation}
Unfortunately, differences between the transformations defined by Eqs.\
(\ref{Def:DeltaTasformation}) and (\ref{Def:dLevinTrDiffOpRep}),
respectively, are greater than they might seem at first sight. Our
convergence analysis of the delta summation of the Euler series is based
on the fact that the truncation error $\mathcal{R}_{n} (\mathcal{E}; z)$
of the Euler series possesses a factorial series representation according
to Eq.\ (\ref{Eq:BorghiANM}). This greatly simplified our analytical
manipulations since both Pochhammer symbols $(n+\beta)_{k-1}$ as well as
factorial series are very convenient objects from the perspective of
finite difference operators. In contrast, powers like $(n+\beta)^{k-1}$
occurring in Eq.\ (\ref{Def:dLevinTrDiffOpRep}) are extremely
inconvenient objects.

On the basis of the available evidence we have to conclude that our
theoretical analysis of the summation of the Euler series cannot be extended to
Levin's $d$ transformation (\ref{Def:dLevinTrDiffOpRep}) and related
variants of Levin's transformation in a straightforward way. One
potential alternative, that could possibly supplement our approach, would
be the use of so-called N\"{o}rlund-Rice integrals for the asymptotic
evaluation of binomial sums $\Delta^{k} F_{n} = \sum_{j=0}^{k} (-1)^{k+j}
\binom{k}{j} F_{n+j}$ as suggested by Flajolet and Sedgewick
\cite{Flajolet/Sedgewick/1995,Flajolet/Sedgewick/2009}. This may well be
a promising idea for future investigations.

\begin{figure}[!ht]
  \centering
  \includegraphics[width=10cm]{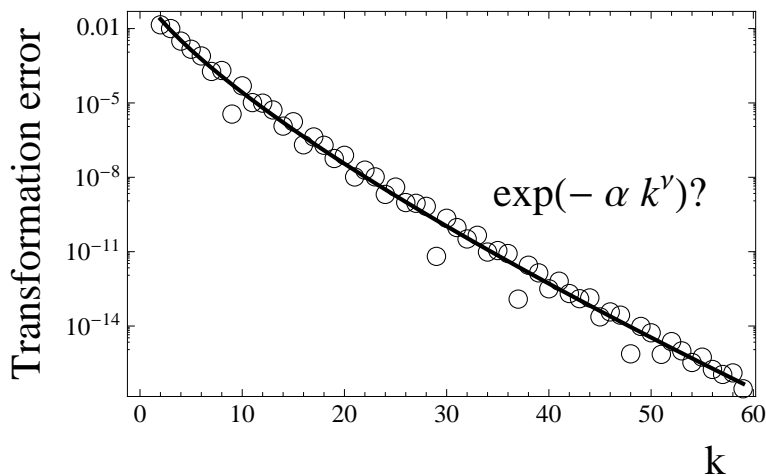}
  \caption{Transformation errors $d^{(0)}_{k} \bigl( 1,\mathcal{E}_n (z)
    \bigr)-\mathcal{E} (z)$ of Levin's $d$ transformation
    (\ref{Def:dLevinTrDiffOpRep}) vs the transformation order $k$ for the
    summation of the Euler series at $z=10$. Open circles represent numerical
    values of $d^{(0)}_{k} \bigl( 1,\mathcal{E}_n (z) \bigr)-\mathcal{E}
    (z)$ evaluated via Eq.\ (\ref{Def:dLevinTrDiffOpRep}). The solid
    curve corresponds to a purely numerical fit of the exponential model
    $d^{(0)}_{k} \bigl( 1,\mathcal{E}_n (z) \bigr)-\mathcal{E} (z)
    \approx A \exp \bigl(-\alpha k^{\nu} \bigr)$. Our results indicate
    $\nu \approx 3/4$.}
  \label{Fig:ComparisonLW}
\end{figure}

In spite of our deplorable lack of theoretical results, we can gain at
least some insight by performing numerical experiments. In Fig.\
\ref{Fig:ComparisonLW} we employ Levin's $d$ transformation
(\ref{Def:dLevinTrDiffOpRep}) for the summation of the Euler series. As
in Figs.\ \ref{Fig:FigDeltaOnEulerz10} and
\ref{Fig:FigDeltaAndPadeOnEulerz10Complex}, we plot in Fig.\ \ref{Fig:ComparisonLW} 
the transformation errors $d_{k}^{(0)} \bigl( 1, \mathcal{E}_{0} (z) \bigr) - \mathcal{E}
(z)$ (open circles) evaluated via Eq.\ (\ref{Def:dLevinTrDiffOpRep}) vs
the transformation order $k$ for $z=10$. The solid curve was obtained as
a purely numerical fit of the exponential model $d^{(0)}_{k} \bigl(
1,\mathcal{E}_n (z) \bigr)-\mathcal{E} (z) \approx A \exp \bigl(-\alpha
k^{\nu} \bigr)$ which resembles our transformation error estimates
(\ref{Eq:BorghiANM.12}) and (\ref{Eq:BorghiANM.14}) obtained for the
delta transformations and diagonal Pad\'{e} approximants.

Several numerical tests carried out for different \emph{positive}
arguments $z$ of the Euler series provide convincing evidence that the
transformation errors $d^{(0)}_{k} \bigl( 1,\mathcal{E}_n (z)
\bigr)-\mathcal{E} (z)$ can for larger transformation orders $k$ and
positive arguments $z$ be modeled by a single exponential $A \exp
\bigl(-\alpha k^{\nu} \bigr)$ and that the exponent $\nu$ is close to
$3/4$. 
Moreover, we also observe in Fig. \ref{Fig:ComparisonLW} some oscillatory
modulations  that could be caused by something resembling the cosine term
in Eq. (\ref{Eq:BorghiANM.12}).

Our results indicate that at least for sufficiently large
transformation orders $k$ the transformation errors $d^{(0)}_{k} \bigl(
1,\mathcal{E}_n (z) \bigr) - \mathcal{E} (z)$ of Levin's $d$
transformation should decay faster than the corresponding transformation
errors $\delta^{(0)}_{k} \bigl(1, \mathcal{E}_{0} (z) \bigr) -
\mathcal{E} (z)$ of the delta transformation, whose decay rate is
according to Eq.\ (\ref{Eq:BorghiANM.12}) characterized by the
exponential $\exp (-9 k^{2/3}/[2 z^{1/3}])$.

Our summation results in Fig.\ \ref{Fig:ComparisonLW} indeed look very
impressive. However, this does not imply that Levin's $d$ transformation
necessarily sums all factorially divergent series more effectively than
the delta transformation. Such a generalization looks tempting but can be badly
misleading. For example, in \cite{Weniger/Cizek/Vinette/1993} the
summation of the factorially divergent Rayleigh-Schr\"{o}dinger
perturbation series for the ground state energy of the quartic anharmonic
oscillator with the help of Levin's $d$ transformation initially seemed
to produce convergent results. But in the case of large transformation
orders the summation results clearly diverged. In the case of the delta
transformation or Pad\'{e} approximants, no divergence was observed in \cite{Weniger/Cizek/Vinette/1993},

The divergence of Levin's transformation was also confirmed in
\cite[Table 2]{Weniger/1992}, where the summations were performed with a
Levin-type transformation that -- depending on the value of a continuous
parameter -- interpolates between Levin's $d$ and the delta
transformation. A similar divergence of Levin's transformation was later
observed by \v{C}\'{\i}\v{z}ek, Zamastil, and Sk\'{a}la \cite[p.\
965]{Cizek/Zamastil/Skala/2003} in the case of the hydrogen atom in an
external magnetic field. More detailed discussions of the divergence of
Levin's transformation can be found in \cite[pp.\ 211 -
216]{Weniger/1994b}, in \cite[pp.\ 7 - 9]{Weniger/2007d}, or in
\cite[pp.\ 57 - 58]{Weniger/2012}.

No completely satisfactory explanation of the divergence
of Levin's $d$ transformation in the case of the anharmonic oscillators
is known. This divergence remains a mystery. Our
inability of explaining this divergence indicates that our current
understanding of the subtleties of summation processes is far from being
satisfactory. There still remains a lot of work to be done.



%
\typeout{==> Section: Conclusions and Outlook}
\section{Conclusions and Outlook}
\label{Sec:Conclusions}
%

The factorially divergent Euler series defined by Eq.\
(\ref{Def:EulerSer}) is an asymptotic series as $z \to 0$ for the Euler
integral defined by Eq.\ (\ref{Def:EulerIntegral}). It is a very
important model problem not only in special function theory, where many
asymptotic expansions diverge factorially, but even more so for
physicists in connection with the ubiquitous factorially divergent
perturbation expansions. The topic of this article is an \emph{in depth}
analysis of the summation of the Euler series to the Euler integral with
the help of Pad\'e approximants and the delta transformation
(\ref{Def:DeltaTasformation}), which is known to be a very powerful
Levin-type sequence transformation. According to experience delta
outperforms the much better known Pad\'{e} approximants in the case of
factorially divergent series.

Our theoretical manipulations produced closed form expressions for the
transformation errors of both Pad\'e and delta. From them, we derived
asymptotic estimates which holds in the case of large transformation
orders. Our estimates clearly shows that the delta transformation is
indeed able to sum the factorially divergent Euler series to the Euler
integral. To the best of our knowledge, our estimate
(\ref{Eq:BorghiANM.12}) is the first explicit theoretical error estimate
for any Levin-type transformation that holds in the limit of infinite
transformation orders. A comparison of our asymptotic transformation
error estimate (\ref{Eq:BorghiANM.12}) for the delta transformation with
the analogous estimate (\ref{Eq:BorghiANM.14}) for diagonal Pad\'{e}
approximants provides the first theoretical explanation of the observed
superiority of the delta transformation in the case of the Euler series
over the much better known Pad\'{e} approximants (compare \cite[Tables
13-1 and 13-2]{Weniger/1989}). We believe that our theoretical results
should give potential users of Levin-type transformations more confidence
in the power and usefulness of these transformations.

In spite of its undeniable success, our approach has obvious
limitations. We were only able to derive our explicit expressions and our
asymptotic estimates because the truncation error
(\ref{Def:TruncErrEuSer}) of the Euler series can be expressed by the
factorial series expansion (\ref{Eq:BorghiANM}). Unfortunately, no
analogous factorial series expansions are known for the truncation errors
of other factorially divergent asymptotic series for special
functions. These unknown factorial series expansions for truncation
errors have to be derived first before one could try to extend our
approach to other special functions. Thus, our article could also be
viewed to be an invitation to mathematicians interested in special
function theory to do something about these missing factorial series
expansions.

Another problem is that our approach cannot be applied to the typical
perturbation expansions of physics because their series coefficients are
usually not known explicitly, but only numerically. For these problems we
need a new approach based on some general principles. Ideal would be
something like the highly developed convergence theory of Pad\'{e}
approximants to Stieltjes series which was briefly reviewed in Section
\ref{Sub:PadeApprox}. Therefore, we actually hope that the results of
this article might constitute a first and probably preliminary step
toward the development of a more general theory of the summation of
factorially divergent series with the help of Levin-type sequence
transformations. But we fear that this will be not easy, and that a lot
of work has to be done before this goal can be achieved.

Our analytical approach also has the obvious shortcoming that it cannot
be extended in a straightforward way to other Levin-type
transformations. As discussed in Section \ref{Sub:LevTr}, we are
currently not able to extend our approach, that was so very successful in
the case of the delta transformation, to the analogous and at least at
first sight very similar $d$ transformation of Levin defined by
Eq. (\ref{Def:dLevinTrDiffOpRep}). It seems that new analytical
techniques have to be developed before we can accomplish anything in this
direction. We already mentioned the so-called N\"{o}rlund-Rice integrals
advocated by Flajolet and Sedgewick
\cite{Flajolet/Sedgewick/1995,Flajolet/Sedgewick/2009}.

Factorially divergent asymptotic and perturbation expansions undoubtedly
constitute challenging numerical problems. However, in quantum physics
perturbation expansions occur whose series coefficients grow roughly like
$(\nu n)!$ with $\nu > 1$, and which therefore constitute even more
challenging problems. Examples are the Rayleigh-Schr\"{o}dinger
perturbation expansions for the energy eigenvalues of the sextic and
octic quantum anharmonic oscillators, whose series coefficients grow
roughly like $(2n)!$ and $(3n)!$, respectively. Pad\'{e} approximants are
not powerful enough to achieve anything substantial in the sextic case,
although they in principle converge, and in the even more challenging
octic case, Graffi and Grecchi \cite{Graffi/Grecchi/1978} showed
rigorously that Pad\'{e} approximants are not able to sum this violently
divergent perturbation expansion. In contrast, the delta transformation
turned out to be successful even for these extremely violently divergent
perturbation expansions \cite{Weniger/1996c,Weniger/Cizek/Vinette/1991,%
  Weniger/Cizek/Vinette/1993}. There can be no doubt that a better
theoretical understanding of the summation of hyper-factorially divergent
expansions would be highly desirable.

Our article most likely provides a definite answer to the problem of the
delta summation of the Euler series and how the delta summation compares 
to Pad\'e summation. 
But otherwise, we would be happy if it could
serve as a starting point for more detailed theoretical investigations on
the summation of divergent series with the help of Levin-type
transformations.

We are fully aware that the topic of our article is mathematical in
nature. However, mathematics is the language of physics, and divergent
series have been and to some extent still are highly controversial. At
the same time, (factorially) divergent series are indispensable in
theoretical physics. We believe that our theoretical results provide
strong evidence that the summation techniques we considered are not only
toys for experimental mathematicians, but safe and reliable numerical
tools. It is a welcome side effect that our results highlight the power
and usefulness of Levin-type transformations which have not yet gained
the recognition they deserve.

\section*{Acknowledgment}

EJW gratefully acknowledges the hospitality of the
Dipartimento di Ingegneria of the Universit\`{a} ``Roma Tre'' where a
part of this work was done.



\newpage


\begin{appendices}
  \typeout{==> Appendix: Stieltjes Functions and Stieltjes Series}
\section{Stieltjes Functions and Stieltjes Series}
\label{App:StieFunStieSer}

A function $f \colon \mathbb{C} \to \mathbb{C}$ is called a
\emph{Stieltjes} function if it can be expressed as a Stieltjes integral:
\begin{equation}
  \label{Def_StieFun}
  f (z) \; = \; \int_{0}^{\infty} \,
  \frac{\mathrm{d} \Phi (t)} {1+zt} \, ,
  \qquad \vert \arg (z) \vert < \pi \, .
\end{equation}
Here $\Phi (t)$ is a bounded, nondecreasing function taking infinitely
many different values on the interval $0 \le t < \infty$. Moreover, the
moment integrals
\begin{equation}
  \label{Def_StieMom}
  \mu_n \; = \; \int_{0}^{\infty} \, t^n \, \mathrm{d} \Phi (t) \, ,
  \qquad n \in \mathbb{N}_0 \, ,
\end{equation}
must be positive and finite for all finite values of $n$.

If we insert the geometric series $1/(1+zt) = \sum_{\nu=0}^{\infty}
(-zt)^{\nu}$ into the integral representation (\ref{Def_StieFun}) and --
ignoring all questions of convergence and legitimacy -- integrate
term-wise using (\ref{Def_StieMom}), we see that a Stieltjes function $f
(z)$ can be expressed at least formally by its corresponding
\emph{Stieltjes} series:
\begin{equation}
  \label{Def_StieSer}
  f (z) \; = \;
  \sum_{\nu=0}^{\infty} \, (-1)^{\nu} \, \mu_{\nu} \, z^{\nu} \, .
\end{equation}
Whether this series converges or diverges depends on the behavior of the
Stieltjes moments $\mu_n$ as $n \to \infty$. 

Because of their highly developed convergence theory in the case of
Pad\'{e} approximants (compare also Section \ref{Sub:PadeApprox}),
Stieltjes series are of considerable importance in the theory of
divergent series. Detailed discussions of Stieltjes series can be found
in books by Bender and Orszag \cite[Chapter 8.6]{Bender/Orszag/1978} and
Baker and Graves-Morris \cite[Chapter 5]{Baker/Graves-Morris/1996}. Other
good sources with a stronger emphasis on mathematical aspects are a
review article \cite{Widder/1938} and a book \cite{Widder/1946} by
Widder.

Let $\{ u_{n} \}_{n=0}^{\infty}$ be a sequence. The Hankel determinants
$H_{k} (u_{n})$ of this sequence are defined as follows (see for example
\cite[pp.\ 78 and 80]{Brezinski/RedivoZaglia/1991a}):
\begin{subequations}
  \label{Def_HankelDeterminant}
  \begin{align}
    \label{Def_HankelDeterminant_a}
    H_{0} (u_{n}) & \; = \; 1 \, , \qquad H_{1} (u_{n}) \; = \; u_{n} \, ,
    && n \in \mathbb{N}_{0} \, ,
    \\
    \label{Def_HankelDeterminant_b}
    H_{k} (u_{n}) & \; = \;
    \begin{vmatrix}
      u_{n} & u_{n+1} & \dots & u_{n+k-1} \\
      u_{n+1} & u_{n+2} & \dots & u_{n+k} \\
      \vdots & \vdots & \ddots & \vdots \\
      u_{n+k-1} & u_{n+k} & \dots & u_{n+2k-2} \\
    \end{vmatrix} \, ,
     && k \in \mathbb{N} \, , \quad n \in \mathbb{N}_{0} \, ,
  \end{align}
\end{subequations}
Hankel determinants play a very important role in the theory of Pad\'{e}
approximants. A necessary condition, that a power series of the type of
(\ref{Def_StieSer}) is indeed a Stieltjes series, is that the Hankel
determinants $H_{k} (\mu_{n})$ of the Stieltes moments
(\ref{Def_StieMom}) of this series are positive for all $k, n \ge 0$ 
\cite[Theorem 5.1.2]{Baker/Graves-Morris/1996}.

Insertion of the partial sum $\sum_{\nu=0}^{n} (-zt)^{\nu} =
\bigl[1-(-zt)^{n+1}\bigr]/\bigl[1+zt\bigr]$ of the geometric series into
Eq. (\ref{Def_StieFun}) shows that a Stieltjes function can be expressed as a
partial sum of the Stieltjes series (\ref{Def_StieSer}) plus a truncation
error term which is also a Stieltjes integral (see for example
\cite[Theorem 13-1]{Weniger/1989}):
\begin{equation}
  \label{Def_StieFunParSum}
  f (z) \; = \; \sum_{\nu=0}^{n} \, (-1)^{\nu} \, \mu_{\nu} \, z^{\nu} 
  \, + \, (-z)^{n+1} \, \int_{0}^{\infty} \, \frac{t^{n+1}\mathrm{d} \Phi
   (t)} {1+zt} \, , \qquad \vert \arg (z) \vert < \pi \, .  
\end{equation}
Moreover, the truncation error term in Eq. (\ref{Def_StieFunParSum})
satisfies -- depending upon the value of $\arg (z)$ -- the following
inequalities (see for example \cite[Theorem 13-2]{Weniger/1989}):
\begin{equation}
  \label{StieSerErrEst}
  \left\vert (-z)^{n+1} \, \int_{0}^{\infty} \, \frac{t^{n+1}
      \mathrm{d} \Phi (t)} {1+zt} \right\vert \; \le \;
  \begin{cases}
    \mu_{n+1} \, \vert z^{n+1} \vert \, ,
    & \qquad \vert \arg (z) \vert \le \pi /2 \, , 
    \rule[-2ex]{0cm}{1em} \\
    \mu_{n+1} \, \vert z^{n+1} \cosec \bigl( \arg (z) \bigr) \vert \, , 
   & \qquad \pi /
    2 < \vert \arg (z) \vert < \pi \, .
  \end{cases}  
\end{equation}


  \typeout{==> Section: Basic Properties of Factorial Series}
\section{Basic Properties of Factorial Series}
\label{App:FactorialSeries}

Let $\Omega (z)$ be a function that vanishes as $z \to + \infty$. A
\emph{factorial series} for $\Omega (z)$ is an expansion of the following
type:
\begin{equation}
  \label{DefFactSer}
  \Omega (z) \; = \; 
  \frac {b_{0}} {z} \, + \, \frac {b_{1} 1!} {z (z+1)}
  \, + \, \frac {b_{2} 2!} {z (z+1) (z+2)} \, + \, \cdots \; = \; 
  \sum_{\nu=0}^{\infty} \frac {b_{\nu} {\nu}!} {(z)_{\nu+1}} \, .
\end{equation}
Here, $(z)_{\nu+1} = \Gamma (z+\nu+1) / \Gamma (z) = z (z+1) \ldots
(z+\nu)$ is a Pochhammer symbol. In general, $\Omega (z)$ will have
simple poles at $z = - m$ with $m \in \mathbb{N}_0$.

The definition of a factorial series according to Eq. (\ref{DefFactSer}) is
typical of the mathematical literature. The separation of the series 
coefficients into a factorial $n!$ and a
reduced coefficient $b_n$ offers some formal advantages.

Factorial series have a long tradition in mathematics. They were
discussed already in Stirling's book \cite{Stirling/1730}, which was
first published in 1730. Recently, a new annotated translation of
Sterling's book was published by Tweddle \cite{Tweddle/2003}, who
remarked that Stirling was not the inventor of factorial
series. Apparently, Stirling became aware of factorial series by the work
of the French mathematician Nicole \cite[p.\
174]{Tweddle/2003}. Nevertheless, Stirling used factorial series
extensively and thus did a lot to popularize them.

In the nineteenth and the early twentieth century, the theory of
factorial series was fully developed by a variety of authors. Fairly
complete surveys of the older literature as well as thorough treatments
of their fundamental properties can be found in classic books on
difference equations by Milne-Thomson
\cite{Milne-Thomson/1981}, Nielsen \cite{Nielsen/1965}, and N\"{o}rlund
\cite{Noerlund/1926,Noerlund/1929,Noerlund/1954}. Factorial series are
also discussed in the books by Knopp \cite{Knopp/1964} and Nielsen
\cite{Nielsen/1909} on infinite series. Additional and in particular more
recent references can be found in \cite{Weniger/2010b}.

In this Appendix, predominantly those properties of factorial series are
discussed that are of importance for an understanding of the convergence
properties of Levin-type transformations. It is extremely easy to apply
higher powers of the finite difference operator $\Delta$ to a factorial
series. In fact, on using
\begin{equation}
  \label{Diff^k_Fact/Poch}
  \Delta^k \frac{n!}{(z)_{n+1}} \; = \; 
  \frac{(-1)^k (n+k)!}{(z)_{n+k+1}} \, , \qquad k,n \in \mathbb{N}_0 \, ,
\end{equation}
which can be proved by complete induction, the application of $\Delta^k$
to a factorial series produces a very simple result:
\begin{equation}
  \label{Diff_k_FactSer}
  \Delta^k \, \Omega (z) \; = \; \sum_{\nu=0}^{\infty} \, 
  \Delta^k \, \frac {b_{\nu} {\nu}!} {(z)_{\nu+1}} 
  \; = \; (-1)^k \, \sum_{\nu=0}^{\infty} \, 
  \frac {b_{\nu} (\nu+k)!} {(z)_{\nu+k+1}} \, .
\end{equation}
This expression can be rewritten as follows:
\begin{subequations}
  \begin{align}
    \Delta^k \, \Omega (z) & \; = \; (-1)^{k} \, \sum_{\kappa=0}^{\infty}
    \, \frac{b_{\kappa}^{(k)} \kappa!}  {(z)_{\kappa+1}} \, ,
    \\
    b_{\kappa}^{(k)} & \; = \;
    \begin{cases}
      0 \, , & \qquad \kappa < k \, , \\
      b_{\kappa-k} \, , & \qquad \kappa \ge k \, .
    \end{cases}
  \end{align}
\end{subequations}
Thus, factorial series play a similar role in the theory of difference
equations as power series in the theory of differential equations. This
explains why factorial series are treated in classic books on finite
differences
\cite{Milne-Thomson/1981,Noerlund/1926,Noerlund/1929,Noerlund/1954}.

Alternative expressions for factorial series can be derived easily. For
example, the beta function is defined by $B (x, y) = \Gamma(x) \Gamma(y)
/ \Gamma(x+y)$ \cite[Eq.\ (5.12.1)]{Olver/Lozier/Boisvert/Clark/2010}.
Accordingly, $B (z, n+1) = n!/(z)_{n+1}$ and the factorial series in
Eq.~(\ref{DefFactSer}) can also be viewed as an expansion in terms of beta
functions (compare for instance \cite[p.\ 288]{Milne-Thomson/1981} or
\cite[p.\ 175]{Paris/Kaminski/2001}):
\begin{equation}
  \label{FactSerBetaExpan}
  \Omega (z) \; = \; \sum_{n=0}^{\infty} \, b_n \, B (z, n+1) \, .
\end{equation}
The beta function possesses the integral representation \cite[Eq.\
(5.12.1)]{Olver/Lozier/Boisvert/Clark/2010}
\begin{equation}
  B (x, y) \; = \; \int_{0}^{1} \, t^{x-1} \, (1-t)^{y-1} \, \mathrm{d} t 
  \, , \qquad \Re (x), \Re (y) > 0 \, ,
\end{equation}
which implies
\begin{equation}
  \label{IntRepFSterm}
  B (z, n+1) \; = \; \frac{n!}{(z)_{n+1}} \; = \;
  \int_{0}^{1} \, t^{z-1} \, (1-t)^{n} \, \mathrm{d} t
  \, , \qquad \Re (z) > 0 \, \quad n \in \mathbb{N}_0 \, .
\end{equation}
Inserting this into (\ref{FactSerBetaExpan}) yields the following
integral representation \cite[Satz I on p.\ 244]{Nielsen/1965},
\begin{subequations}
  \label{FS_IntRep}
  \begin{align}
    \label{FS_IntRep_a}
    \Omega (z) & \; = \; \int_{0}^{1} \, t^{z-1} \, \varphi (t) \,
    \mathrm{d} t \, , \qquad \Re (z) > 0 \, ,
    \\
    \label{FS_IntRep_b}
    \varphi (t) & \; = \; \sum_{n=0}^{\infty} \, b_n \, (1-t)^n \, .
  \end{align}
\end{subequations}
Frequently, the properties of $\Omega (z)$ can be studied more easily via
this integral representation than via the defining factorial series
(\ref{DefFactSer}) (see for example \cite[Chapter XVII]{Nielsen/1965}).

The integral representation (\ref{FS_IntRep}) can also be used for an
alternative derivation of (\ref{Diff_k_FactSer}). For that purpose, we
use
\begin{equation}
  \Delta \, t^{z} \; = \;t^{z+1}\,-\,t^z\; = \; (t-1) \, t^{z}\,,
\end{equation}
or equivalently
\begin{equation}
  \Delta^{k} \, t^{z} \; = \; (-1)^{k} \, (1-t)^{k} \, t^{z} \, .
\end{equation}
Inserting this into the integral representation (\ref{FS_IntRep}) yields
with the help of (\ref{IntRepFSterm}):
\begin{align}
  \Delta^{k} \, \Omega (z) & \; = \; (-1)^{k} \, \sum_{n=0}^{\infty} \,
  b_{n} \, \int_{0}^{1} \, t^{z-1} \, (1-t)^{k+n} \, \mathrm{d} t
  \\
  & \; = \; \sum_{n=0}^{\infty} \, \frac{b_{n} (k+n)!}{(z)_{k+n+1}} \, .
\end{align}

  
\end{appendices}

%
%
\bibliographystyle{WithTitles_NB}
%

%
%
\newpage \addcontentsline{toc}{section}{Bibliography}
\bibliography{BorghiWeniger}
%

\end{document}